\documentclass[]{aa}  

\usepackage{natbib}
\bibpunct{(}{)}{;}{a}{}{,}
\usepackage{graphicx}
\usepackage{revsymb}	
\usepackage{amsmath}	
\usepackage[usenames]{color}	
\usepackage{txfonts}
\usepackage{amssymb}
\usepackage{color}
\usepackage{adjustbox}
\usepackage{geometry} 
\usepackage[parfill]{parskip} 
\usepackage{amssymb}
\usepackage{slantsc}
\usepackage{array}
\usepackage{url}
\usepackage{amsfonts}
\usepackage[colorlinks=true,linkcolor=black, urlcolor=black, citecolor=black]{hyperref}
\usepackage{sidecap}
\usepackage{graphicx}
\usepackage{xcolor}
\usepackage{nicefrac}
\usepackage{subfigure} 
\usepackage{epsfig}
\colorlet{rouge}{red!70!darkgray}
\usepackage{booktabs, caption} 

\begin{document}

\title{Revisiting Kepler-444}
\subtitle{II. Rotational, orbital, and high-energy fluxes evolution of the system}
\author{C. Pezzotti\inst{1} \and P. Eggenberger\inst{1} \and G. Buldgen\inst{1} \and G. Meynet\inst{1} \and {V. Bourrier}\inst{1} \and C. Mordasini \inst{2} }
\institute{Observatoire de Genève, Université de Genève, Chemin Pegasi 51, CH$-$1290 Sauverny, Suisse \and Physikalisches Institut, University of Bern, Gesellschaftsstrasse 6, CH-3012 Bern, Switzerland}
\date{April, 2021}
\abstract{Kepler-444 is one of the oldest planetary systems known thus far. Its peculiar configuration consisting of five sub-Earth-sized planets orbiting the companion to a binary stellar system makes its early history puzzling. Moreover, observations of HI-Ly$\rm \alpha$ variations raise many questions about the potential presence of escaping atmospheres today.}
{We aim to study the orbital evolution of Kepler-444-d and Kepler-444-e and the impact of atmospheric evaporation on Kepler-444-e.}
{Rotating stellar models of Kepler-444-A were computed with the Geneva stellar evolution code and coupled to an orbital evolution code, accounting for the effects of dynamical, equilibrium tides and atmospheric evaporation. The impacts of multiple stellar rotational histories and extreme ultraviolet (XUV) luminosity evolutionary tracks are explored.}
{Using detailed rotating stellar models able to reproduce the rotation rate of Kepler-444-A, we find that its observed rotation rate is perfectly in line with what is expected for this old K0-type star, indicating that there is no reason for it to be exceptionally active as would be required to explain the observed HI-Ly$\rm \alpha$ variations from a stellar origin. We show that given the low planetary mass ($\rm \sim 0.03~M_{\oplus}$) and relatively large orbital distance ($\rm \sim 0.06$ AU) of Kepler-444-d and e, dynamical tides negligibly affect their orbits, regardless of the stellar rotational history considered. We point out instead how remarkable the impact is of the stellar rotational history on the estimation of the lifetime mass loss for Kepler-444-e. We show that, even in the case of an extremely slow rotating star, it seems unlikely that such a planet could retain a fraction of the initial water-ice content if we assume that it formed with a Ganymede-like composition.
}{}

\maketitle

\section{Introduction}

The great variety of planetary systems discovered in the last two decades, with populations of gaseous, rocky planets orbiting different type of stars \citep[e.g.][]{Perryman2011, Howard2012}, has made it necessary to address several open questions about the origin and evolution of these systems as a whole. In this context, a natural synergy has been established between the asteroseismology and exoplanetology communities \citep[e.g.][]{Christensen2010, Batalha2011, Huber2013a, Huber2013b, Huber2018, Campante2018}, providing a highly accurate characterisation of the host star, which leads to a more accurate determination of the planetary parameters \citep[e.g.][]{Reese2012,Buldgen2015}. Developing more refined and sophisticated stellar models, in particular regarding the rotational properties and the exchange of angular momentum (AM) between the host star and the planetary orbits, is crucial to better constrain the evolutionary history of the systems.

In this framework, we aim to revisit the Kepler-444 system. We have provided a detailed seismic characterisation in \citet[][Paper I hereafter]{2019BuldgenK}, leading to new determinations for the mass, radius, and age of the host star ($\rm  0.754 \pm 0.030 \, M_{\odot}$, $\rm 0.753 \pm 0.010 \, R_{\odot}$,$\rm 11.0 \pm 0.8\,$Gyr), and hence new estimates for the mass of the Kepler-444-d and e planets. These values were found to be in good agreement with previous determinations obtained by \citet{Campante2015}. Here we would like to take a step further, studying the orbital evolution of Kepler-444-d and Kepler-444-e and the eventual impact of atmospheric evaporation.

Kepler-444 is one of the oldest planetary system known thus far. This is a rather peculiar system, composed of five sub-Earth-sized planets orbiting the primary star Kepler-444-A, within $\rm 0.1$ AU and $\rm P_{orb} < 10$ days, along with two spatially unresolved M dwarfs (Kepler-444-BC) at a projected distance of 66 AU \citep{Lillo-Box2014,Dupuy2016}. The highly eccentric stellar orbit ($\rm e =  0.864 \pm 0.023$) brings the M dwarf binary companion very close to the planetary system every $\rm \sim 460$ yr \citep{Campante2015, Mills2017}, within $\rm 5^{+0.9}_{-1.0}$ AU. This likely primordial orbital configuration is expected to have strongly impacted the protoplanetary disc, truncating it at $\rm \approx 2$ AU and severely depleting it of the solid material from which the total $\rm \approx 1.5  ~ M_{\oplus}$ planetary mass would have formed \citep{Dupuy2016}. This particular configuration makes the early history of the system puzzling since the formation and migration of planets in such a truncated disc are not well understood \citep{Papaloizou2016,Mills2017}. Some studies suggest that such tightly packed planetary systems would have formed locally; however, the estimated densities for the two outermost planets Kepler-444-d and e, being in between a Mercury-like rock/iron ratio and water world composition, would accredit the hypothesis of a formation beyond the snowline  \citep{Mills2017, 2019BuldgenK}.

\citet{Bourrier2017} detected significant variations in the Kepler-444-A Ly-$\rm \alpha$ signal at different observational epochs. They observed flux variations of the order of $\rm 20\%$ and $\rm 40\%$ during the transiting and non-transiting periods, respectively. These variations could be related to stellar activity, explaining the signal detection in and out of transit. However, \citet{Bourrier2017} argue that this scenario is unlikely given the quietness of this old main-sequence star. Instead, they propose that the $\rm \sim 20\%$ variation could be due to the presence of escaping hydrogen-rich atmospheres from the two outermost planets (Kepler-444-e and Kepler-444-f) and the $\rm \sim 40\%$ variation could result from the presence of an undetected Kepler-444-g planet as of yet at a larger orbital distance. If the planetary origin of the signal is confirmed, \cite{Bourrier2017} propose that Kepler-444-e and f could be examples of ocean planets. They would have formed beyond the snowline and migrated to the current orbital distance, retaining a substantial amount of water ice to replenish the escaping atmosphere along the $\rm \sim 11$ Gyr of evolution \citep{Bourrier2017}.

In this study, we use the comprehensive asteroseismic characterisation of Kepler-444-A performed in Paper I to provide a detailed modelling of this planetary system, which accounts for the rotational evolution of the host star together with the impact of tides and X-ray and extreme ultraviolet (EUV) fluxes on the planets. We use the Geneva stellar evolution code \citep{Eggenberger2008} that follows the rotational evolution of the star by computing the internal angular momentum (AM) transport through shear instability, meridional circulation, and magnetic instabilities \citep{Eggenberger2010, Eggenberger2010b, Eggenberger2019a}. These stellar models\footnote{The rotating stellar models of Kepler-444-A and the relative XUV fluxes computed with the Geneva stellar evolution code are freely available on \url{https://www.unige.ch/sciences/astro/evolution/fr/recherche/geneva-grids-stellar-evolution-models/}.} are then coupled to the orbital evolution code \citep{Privitera2016A, Privitera2016B, Privitera2016c, Meynet2017, Rao2018} to study the impact of dynamical and equilibrium tides on the architecture of the system. We also study the impact of atmospheric evaporation on Kepler-444-e, using the prescription for the hydrodynamical escape in the energy-limited regime \citep{Lecavelier2007,Erkaev2007}. We investigate the impact of different rotational histories, XUV-luminosity prescriptions, and heating efficiencies on the estimation of the initial planetary mass.

\section{Method \& implemented physics}

In order to characterise the evolution of a planetary system, providing robust and precise stellar parameters for the host star through asteroseismic modelling is crucial. Accurate parameters of the host star (in particular, the mass, radius, and age) are indeed needed to better understand the global evolution of the planetary system. In the present study, the stellar parameters determined in Paper I were used as key inputs and constraints to compute rotating stellar models of Kepler-444. The computation of the evolution of the host star was then coupled to the computation of the evolution of the orbits of the planets to obtain a global view of the evolution of the system. In what follows, we first briefly recall the physics implemented in the stellar and orbital evolution codes.

\subsection{Geneva stellar evolution code}
\label{sect:genec}

Stellar models are computed by using a version of the Geneva stellar evolution code (GENEC) specifically optimised for the evolution of low- and intermediate-mass stars. The physics implemented in the code is described in detail in \cite{Eggenberger2008}; here, we only briefly recall the fundamental properties that are of interest for the present study. Rotational effects are included with the hypothesis of shellular rotation \citep{Zahn1992}. The internal AM transport was followed simultaneously with the evolution of the star by accounting for meridional currents, transport by the shear instability, and transport by magnetic fields. The transport by magnetic fields was taken into account in the context of the Tayler-Spruit dynamo \citep{Spruit2002}. While correctly accounting for the approximately uniform rotation in the radiative interior of the Sun, we recall that the exact description of this process as well as its viability under realistic stellar conditions is still a matter of debate \citep{bra06,zah07,bra17,ful19}. This of course illustrates that the general modelling of the impact of magnetic fields on internal stellar properties remains an open question.
The equation describing the transport of AM is then given by the following:

\begin{equation}
\rm \rho \dfrac{d}{dt} \left( r^{2} \Omega \right)_{M_{r}} = \dfrac{1}{5 r^{2}} \dfrac{\partial}{\partial r} \left( \rho r^{4} \Omega U(r) \right) + \dfrac{1}{r^{2}} \dfrac{\partial}{\partial r} \left( \rho \left( D_{shear} + \nu_{TS} \right) r^{4} \dfrac{\partial \Omega}{\partial r} \right),
\end{equation}

with $r$ being the radius, and $\rho(r)$ and $\Omega(r)$ being the mean density and angular rotation on an isobar, respectively. The velocity $U(r)$ corresponds to the radial component of the meridional circulation velocity in the radial direction and $D_{ \rm shear}$ is the diffusion coefficient for AM transport by the shear instability \citep{Eggenberger2010c}. The viscosity term $\nu_{ \rm TS}$ was included to account for the AM transport by magnetic fields, which only operates when the radial differential rotation is larger than a given threshold \citep{Spruit2002}. Atomic diffusion was included and diffusion coefficients were computed following the prescription by \citet{Paquette1986}. A solar calibrated value of the mixing-length parameter was used for convection. In addition to internal AM transports, the braking of the stellar surface rotation due to magnetised winds was taken into account by using the braking law of \citet{Matt2015, Matt2019}, which describes the global angular momentum loss of Sun-like stars, defined in this case as stars with masses ranging between about $\rm 0.5$ and $\rm 1.3~M_{\odot}$, with outer convective envelopes that are magnetically active \citep{Matt2015,Matt2019}. The equations for the torque read as follows: 

\begin{equation}
\rm \dfrac{dJ}{dt} =
\begin{cases}
\rm -T_{\odot} \left(\dfrac{R}{R_{\odot}} \right)^{3.1} \left( \dfrac{M}{M_{\odot}} \right)^{0.5} \left(\dfrac{\tau_{cz}}{\tau_{cz \odot}} \right)^{p} \left(  \rm \dfrac{\Omega}{\Omega_{\odot}} \right)^{p+1} &, \rm \text{if} ~ \left( Ro > Ro_{\odot}/\chi \right),\\
\rm -T_{\odot} \left(\dfrac{R}{R_{\odot}} \right)^{3.1} \left( \dfrac{M}{M_{\odot}} \right)^{0.5} \chi^{p} \left( \dfrac{\Omega}{\Omega_{\odot}} \right) & , \rm \text{if} ~ \left( Ro \leq Ro_{\odot}/\chi \right) . 
\end{cases}
\end{equation}

We note that $\rm R_{\odot}$ and $\rm M_{\odot}$ are the radius and the mass of the Sun, $\rm R$ and $\rm M$ are the radius and the mass of the stellar model, $\rm \tau_{cz}$ is the convective turnover timescale, and $\rm Ro$ is the Rossby number, defined as the ratio between the stellar rotational period and the convective turnover timescale ($\rm Ro= P_{\star}/\tau_{cz}$). The quantity $\rm \chi \equiv Ro_{\odot} / Ro_{sat}$ indicates the critical rotation rate for stars with a given $\rm \tau_{cz}/\tau_{cz_{\odot}}$, defining the transition from the saturated to unsaturated regime. In this work, $\rm \chi$ is considered equal to $\rm 10$ or $\rm 13$ depending on the relation adopted between the Rossby number and the predicted X-ray luminosity (see Sect.~\ref{sect:orbit}). The exponent $\rm p$ is considered equal to $\rm 2.3$ and the constant $\rm T_{\odot}$ was calibrated in order to reproduce the solar surface rotation rate \citep{Eggenberger2019a}. We recall that, once this calibration is performed, this braking law is expected to correctly predict the global angular momentum loss of stars with different masses \citep{Matt2015}. This is due to the fact that the effects of a change in mass on the physical properties of convective envelopes are consistently taken into account in this expression through the computation of the convective turnover timescales. In addition to the internal transport of AM and to the torque due to magnetised winds, the exchanges of AM between the star and the planetary orbits due to equilibrium and dynamical tides are taken into account as described in Sect.~\ref{sect:orbit}. 

Solar models computed with the physics described above have been proven to be in very good agreement with helioseismic data in reproducing the approximately flat rotation profile in the solar radiative zone, thanks to the contribution of magnetic instabilities in the transport of AM \citep{Eggenberger2005, Eggenberger2019a}. Moreover, the same models are also able to correctly reproduce the distribution of surface rotation rates observed for solar-like stars in open clusters \citep{Eggenberger2019a}. The capability of these models to reproduce surface and internal rotational properties of solar-like stars well is of key importance for modelling the rotational evolution of a planet host star and to study its impact on the global evolution of an exoplanetary system.

\subsection{Orbital evolution code}
\label{sect:orbit}

The evolution of exoplanetary systems was computed using the orbital evolution code of \citet{Privitera2016A,Privitera2016B, Privitera2016c}, \citet{Meynet2017} and \citet{Rao2018}, considering the case of a planet on a circular, coplanar orbit around its host star. A full model of the host star was computed as explained above, giving as initial inputs the stellar parameters obtained from the asteroseismic characterisation of the system, going from the pre-main sequence (PMS) phase to the end of the main sequence (MS) or up to the red giant branch (RGB) tip, depending on the physical case we are interested in. Stellar models were then coupled with the orbital evolution code in which we gave the initial orbital distance, the mass of the planet, and the initial angular velocity of the host star as input. In our simulations, we did not account for orbital migration mechanisms occurring across the protoplanetary discs; we did indeed start our computations once the protoplanetary disc had completely disappeared and followed the eventual, subsequent migration of the planets related to the impact of tidal dissipation in the host star convective envelope.
We recall that in our study, we do not consider possible mutual interactions among planets. The physics of the code is described in \citet{Rao2018}. In the following, we briefly recall the fundamental equations. The total change of the orbital distance is as follows:

\begin{equation}
\rm \left(\dot{a}/a \right) = - \dfrac{\dot{M}_{\star} +\dot{M}_{pl}}{ M_{\star} + M_{pl}} - \dfrac{2}{M_{pl} v_{pl}} \left[ F_{fri} + F_{gra} \right] + \left( \dot{a}/a\right)_{t},
\end{equation}

where $\rm \dot{M}_{\star} = - \dot{M}_{loss}$ is the stellar mass loss rate, $\rm M_{pl}$ and $\rm \dot{M}_{pl}$ are the planetary mass and the rate of change of the planetary mass, and $\rm v_{pl}$ is the planetary orbital velocity. We note that $\rm F_{fri}$ and $\rm F_{gra}$ are the frictional and gravitational drag forces, respectively, whose expressions are the same as in \citet{Villaver2009}, \citet{Mustill2012} and \citet{Villaver2014}. The effect of dynamical and equilibrium tides is included in the term $\rm \left( \dot{a}/a \right)_{t} $. The expression for the equilibrium tides \citep{Zahn1966, Alexander1976, Zahn1977,  Zahn1989, LS1984b, Villaver2009, MV2012, Villaver2014} is the same as in \citet{Privitera2016B}. The effect of equilibrium tides is only taken into account when an external convective envelope is present. The expression for equilibrium tides is as follows:

\begin{equation}
\rm \left( \dot{a}/a \right)_{eq} = \frac{f}{\tau_{cz}} \frac{M_{env}}{M_{\star}} q(1+q) \left( \frac{R_{star}}{a} \right)^8 \left( \frac{\Omega_{\star}}{\omega_{pl}} - 1\right),
\end{equation}

where $\rm f$ is a numerical factor obtained from integrating the viscous dissipation of the tidal energy across the convective zone \citep{Villaver2009}, $\rm M_{env}$ is the mass of the convective envelope, $\rm q$ is the ratio between the mass of the planet and the one of the star ($\rm q = M_{pl}/M_{\star}$), $\rm \Omega_{\star}$ is the angular velocity at the stellar surface, and $\rm \omega_{pl} = 2\pi/ P_{orb}$ is the orbital angular velocity of the planet; $\rm \tau_{cz} $ is the convective turnover timescale already introduced in the preceding section. 

In addition to equilibrium tides, we also accounted for the impact of dynamical tides, namely for the frequency-average tidal dissipation of inertial waves excited in the convective envelope of the rotating star, whose restoring force is the Coriolis force \citep{BolmontMathis2016,Gallet2017,Bolmont2017,Benbakoura2019}.
Their effect is accounted for when the condition $\rm \omega_{pl} < 2~\Omega_{\star}$ is satisfied since a planetary companion in a circular-coplanar orbit around a uniformly rotating host star is able to excite inertial waves when the orbital frequency is lower than twice the stellar angular rotation \citep{Ogilvie2007}. We use the following expression for dynamical tides given by \citet{Ogilvie2013} and \citet{Mathis2015}:

\begin{equation}
\rm \left( \dot{a}/a \right)_{dyn} = \left( \dfrac{9}{2Q^{\prime}_{d}} \right)q \omega_{pl} \left( \frac{R_{\star}}{a} \right)^5 \dfrac{(\Omega_{\star} - \omega_{pl})}{\mid \Omega_{\star} - \omega_{pl}\mid},
\end{equation}

with $\rm Q^{\prime}_{d} = 3/(2D_{\omega})$ and  $\rm D_{\omega} = D_{0\omega}D_{1\omega}D_{2\omega}^{-2}$. The `D' terms are as follows:

\begin{equation}
\begin{cases}
\rm D_{0\omega} = \dfrac{100\pi}{63} \epsilon^{2} \dfrac{\alpha^5}{1 - \alpha^5} (1 - \gamma)^2,\\
\rm D_{1\omega} = (1 - \alpha)^4 \left( 1 + 2\alpha + 3\alpha^2 + \frac{3}{2} \alpha^3 \right)^2 ,\\
\rm D_{2\omega} = 1 + \frac{3}{2}\gamma + \frac{5}{2 \gamma}\left( 1 + \frac{\gamma}{2} - \frac{3 \gamma^2}{2} \right) \alpha^3 - \frac{9}{4}\left(1 - \gamma\right)\alpha^5  ,
\end{cases}
\end{equation}

where $\rm \alpha = R_{c}/R_{\star}$, $\rm \beta = M_{c}/M_{\star}$, $\rm \gamma = \dfrac{\alpha^3 (1 - \beta)}{\beta (1 - \alpha^3)}$, and $\rm \epsilon = \dfrac{\Omega_{\star}}{\sqrt{\dfrac{GM_{\star}}{R_{\star}^3}}}$. We note that $\rm M_{c}$ and $\rm R_{c}$ are the mass and the radius of the radiative core, considered as the region of the star below the base of the convective envelope. The term $\rm D_{\omega}$ was computed in \citet{Ogilvie2013} as the frequency-averaged tidal dissipation. This dissipation is referred to as the kinetic energy of the wave-like solution, which was obtained from the dynamical equations system for the stellar envelope \citep{Ogilvie2013,Mathis2015}.  

Tides become inefficient when the planetary distance is equal to the corotation radius, namely when the orbital frequency of the planet is equal to the stellar surface angular rotation ($\rm \omega_{pl} = \Omega_{\star}$). The effect that tides have is to widen the orbit when the planet is beyond the corotation radius and to shrink it when the planet is inside.

We also consider the evolution and impact of the stellar XUV flux on the exoplanetary atmosphere. Following the work by \citet{Tu2015}, we first used the formula for the computation of the XUV flux of \citet{Wright2011}. This prescription gives the ratio of the X-ray luminosity with respect to the bolometric luminosity of the star ($\rm R_{X} = L_{X}/L_{\star} $) in saturated and unsaturated regimes as a function of the Rossby number. The value of the Rossby number establishes the transition from one regime to the other (see Sect.~\ref{sect:genec}): 

\begin{equation}
\begin{cases}
\rm R_X  = 10^{-3.13} \times \left( \dfrac{Ro}{Ro_{sat}} \right)^{-\beta} \qquad &, \rm \text{if} \qquad Ro >  Ro_{\odot}/\chi\\
\rm R_X = 10^{-3.13} \qquad &, \rm \text{if} \qquad Ro \leq  Ro_{\odot}/\chi.\\
\end{cases}
\label{eq:Tu2015}
\end{equation}

A value of $\chi=10$ was used with this expression for the transition from a saturated to unsaturated regime. In the saturated regime, a constant value $R_{X}=10^{-3.13}$ was adopted as given by \citet{Wright2011}. In the unsaturated regime, a solar-calibrated value was used for the exponent $\rm \beta$ with a value of -3.25. We note that this value slightly differs from the value of -2.7 reported by \citet{Wright2011}, which reflects differences in the computation of the Rossby numbers.\ In the present case, the Rossby numbers were directly computed from the properties of our rotating models.
As in \citet{Tu2015}, we also used the prescription of \citet{SanzForcada2011} to obtain the corresponding EUV luminosity:

\begin{equation}
\rm \log_{10}( L_{\rm EUV}) = 4.8 + 0.86 \log_{10} (L_{\rm X}) .
\label{eq:SF}
\end{equation}

The expressions given by Eqs.~\ref{eq:Tu2015} and \ref{eq:SF} were used by default in the present study to compute the X-ray and EUV luminosities from the structural and rotational properties of our models. However, more recent studies suggest different prescriptions for these relations, in particular regarding the hypothesis of a constant value of the $R_X$ ratio in the saturated regime \citep[e.g.][]{Reiners2014,Johnstone2020}. Following the work by \citet{Johnstone2020}, we then also included a more recent prescription defining the evolution of the X-ray luminosity as a function of the Rossby number:

\begin{equation}
\begin{cases}
\rm R_X  =  C_{1}~ Ro^{\beta_{1}}\qquad &, \rm \text{if} \qquad Ro > Ro_{\odot}/\chi\\
\rm R_X = C_{2} ~Ro^{\beta_{2}}\qquad &, \rm \text{if} \qquad Ro \leq Ro_{\odot}/\chi.\\
\end{cases}
\label{eq:J2020}
\end{equation}

The main difference between this expression and Eq.~\ref{eq:Tu2015} is related to the fact that a dependence on the Rossby number is also taken into account in the saturated regime. As for the case of the expression of \citet{Wright2011}, the values of the different coefficients entering Eq.~\ref{eq:J2020} cannot be directly taken from the work of \citet{Johnstone2020} due to differences in the way Rossby numbers are computed. Based on the work of \citet{Johnstone2020}, we then determined the constants and exponents of the broken power law of Eq.~\ref{eq:J2020} by doing a linear fit of the $\rm R_X - Ro$ distribution of stars taken from the catalogue of \citet{Wright2011}. A value of $\chi=13$ is then obtained for the transition between the saturated and unsaturated regime with the following values for the coefficients: $\rm C_{1} = 6.27 \times 10^{-6}$, $\rm C_{2} = 3.71 \times 10^{-4}$, $\rm \beta_{1} = -2.275$, and $\rm \beta_{2} = -0.177$. The non-zero value of $\beta_{2}$ confirms that a small dependence on the rotation rate is favoured over a constant value of $R_X$ in the saturated regime, which is in good agreement with the results of \citet{Johnstone2020}. Moreover, the value determined for $\rm \beta_{1}$ indicates that a steeper slope is obtained in the unsaturated regime when the relation is calibrated to reproduce the solar values as done previously in Eq.~\ref{eq:Tu2015}. As a consequence, using Eq.~\ref{eq:J2020} instead of Eq.~\ref{eq:Tu2015} leads to higher values for the X-ray luminosity for old main-sequence solar-type stars (see discussion in Sect.~\ref{rot_hist_XUV} for the impact of these different prescriptions in the specific case of Kepler-444).

In addition to suggesting a new $R_X$ - Ro relation, \citet{Johnstone2020} also reconsidered the relation between the EUV and X-ray fluxes. Consequently, when Eq.~\ref{eq:J2020} is adopted for the computation of the X-ray luminosity, the corresponding equations for the EUV fluxes for the hard (10-36 nm) and soft (36-92 nm) components, respectively, are also the same as in \citet{Johnstone2020} and read as:
\begin{align} \label{eq:J2020_  euv}
\rm \log_{10} (F_{EUV,1}) &= \rm 2.04 + 0.681 \log_{10} (F_{X}), \nonumber \\
\rm \log_{10} (F_{EUV,2}) &= \rm -0.341 + 0.920 \log_{10}(F_{EUV,1}) .
\end{align}

\section{Rotational and orbital evolution}

\subsection{Rotational history of the system}
\label{rothist}

\begin{figure}
\includegraphics[width=\linewidth]{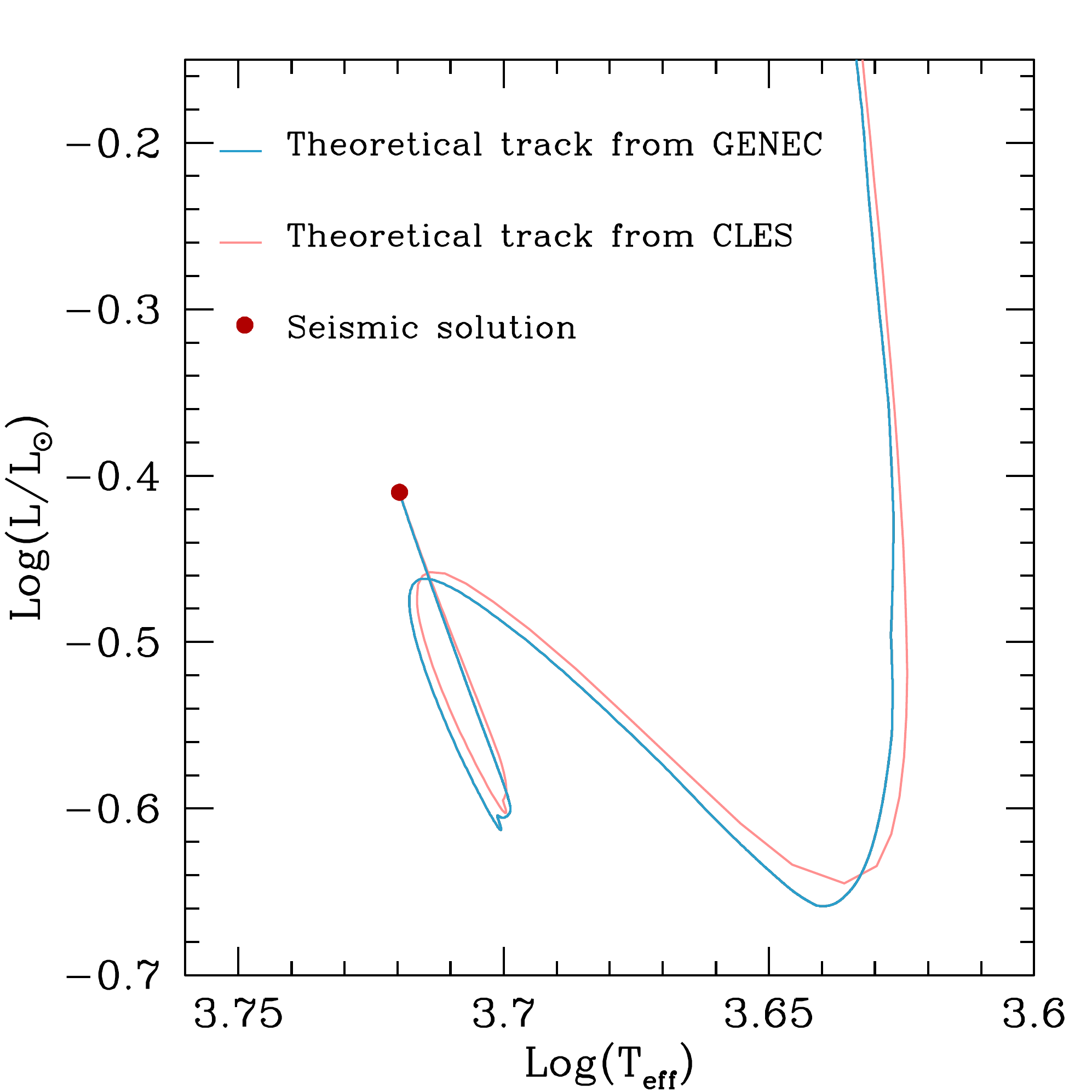} 
\caption{\small{Evolutionary tracks in the HR diagram obtained with the GENEC (in blue) and CLES (in salmon) evolution codes together with the location of the optimal asteroseismic solution (red dot). The tracks show the evolution from the PMS to part of the MS phase. The GENEC track in particular corresponds to the moderate rotation case.}}
\label{dhr}
\end{figure}

Rotating models of Kepler-444 have been computed with the GENEC code using the input physics described in the previous section. The initial parameters of these models are taken from Paper I in order to reproduce the optimal asteroseismic solution. The rotational history of Kepler-444-A being unknown, we computed models representative of an initially slow, moderate, and fast rotating star by using different values for the initial rotation velocity and disc lifetimes during the PMS. The choice of these values was determined from surface rotation rates observed for solar-type stars in open clusters of various ages as was done in \citet{Eggenberger2019a}. Models corresponding to slow and moderate rotators were then computed with an initial rotation rate $\Omega_{\rm ini}$ of 3.2, 5, and 18\,$\Omega_\odot$, respectively. Figure~\ref{dhr} shows the evolution in the HR diagram of the moderate rotating case (blue line) together with the evolutionary track corresponding to the non-rotating model computed with the CLES code in Paper I (salmon line). The location in the HR diagram of the optimal asteroseismic solution is denoted by a red dot.

\begin{figure}[t]
\centering
\subfigure{\includegraphics[width=0.45\textwidth]{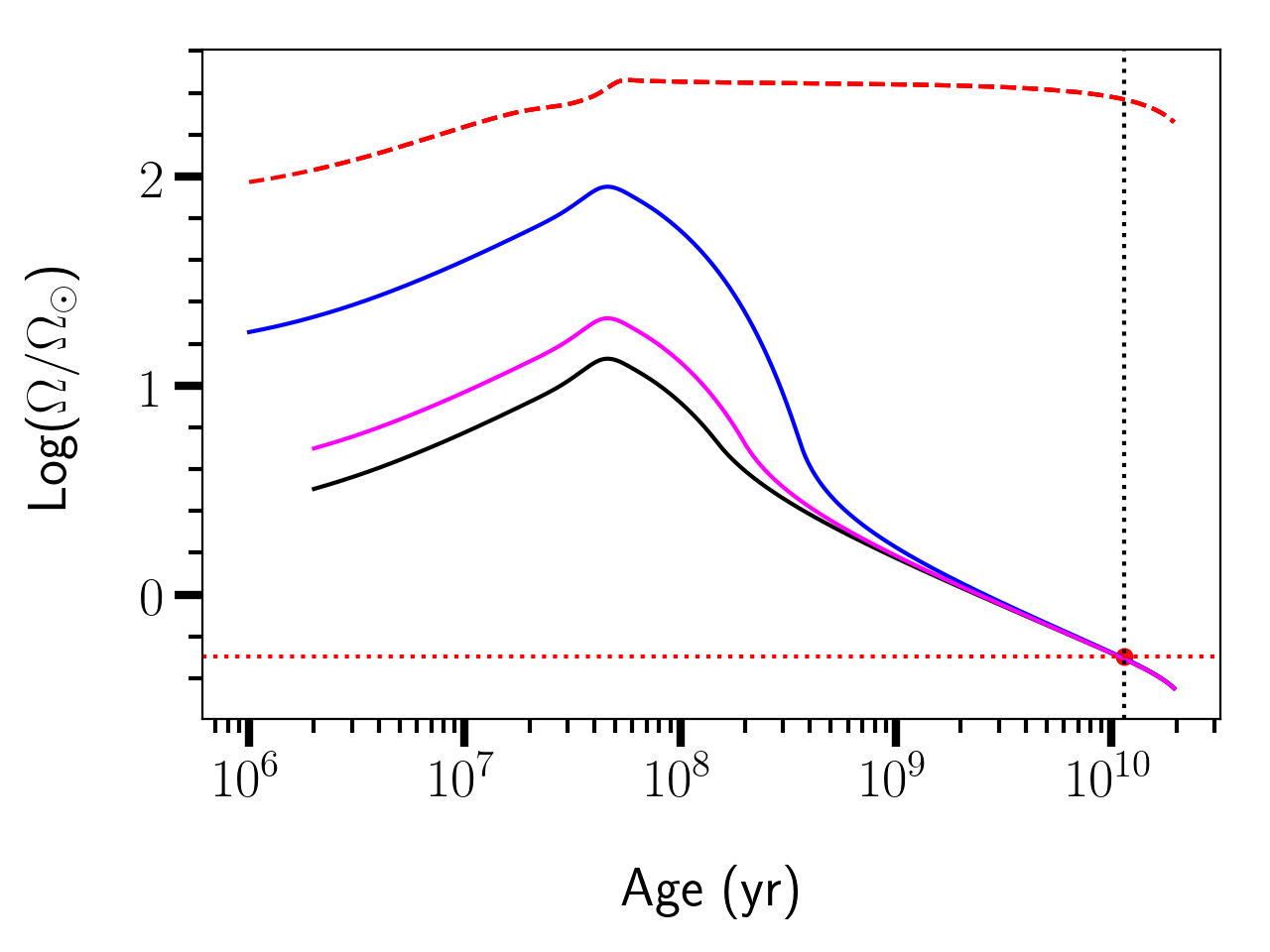}} 
\caption*{}
\subfigure{\includegraphics[width=0.47\textwidth]{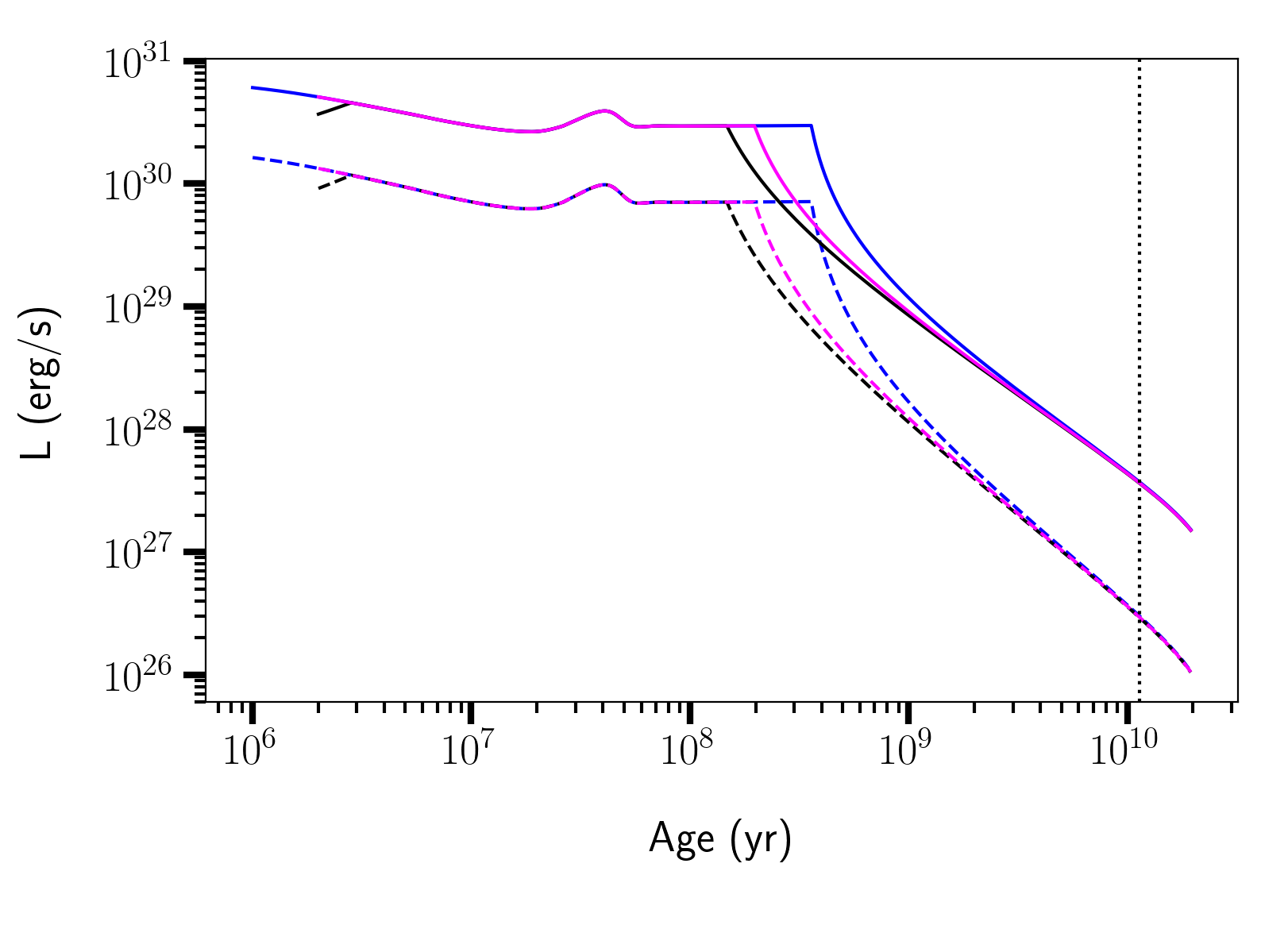}}
\caption*{}
\caption{\footnotesize{\emph{Top panel:} Surface angular velocity evolution for Kepler-444 in the case of a fast rotator ($\rm \Omega_{ini} = 18 \times \Omega_{\odot}$, solid blue line), a medium rotator ($\rm \Omega_{ini} = 5 \times \Omega_{\odot}$, solid magenta line), and a slow rotator ($\rm \Omega_{ini} = 3.2 \times \Omega_{\odot}$, solid black line). The red dot represents the surface angular velocity of Kepler-444-A at the age of the system (11 Gyr). The red-dashed line shows the critical velocity limit. The vertical black-dotted line indicates the age of the system, while the horizontal red-dotted line shows the value of the surface rotation at the age of the system. \emph{Bottom panel:} Evolution of the EUV (solid lines) and X-ray luminosities (dashed lines) computed according to Eqs.~\ref{eq:Tu2015} and \ref{eq:SF} relative to the different rotational histories considered. The vertical black-dotted line indicates the age of the system.}}
\label{rotational_history}
\end{figure}

In the top panel of Fig. \ref{rotational_history}, we show the evolution of the surface rotation as a function of the age for the slow, moderate, and fast rotating cases. After the disc-locking phase, during which the surface angular velocity remains constant, the star experiences a spin up due to the rapid contraction on the PMS phase. The surface rotation reaches a peak at approximately $\rm 48$ Myr, then it starts to decrease due to the braking of the surface induced by magnetised stellar winds. Independently from the initial value of $\rm \Omega_{ini}$, all the surface rotation tracks then overlap after $\rm \sim 10^{3}$ Myr, converging and reproducing the surface rotation of the star well at the age of the system: $\rm \Omega_{K444} = 1.47 \times 10^{-6}$ Hz ($\rm P_{rot} = 49.40 \pm 6.04$ d, \citet{Mazeh2015}). We thus find that rotating models able to correctly reproduce the rotational properties of the Sun and the evolution of surface rotation rates of solar-type stars in open clusters predict a surface rotation rate for Kepler-444-A that is in perfect agreement with the observed value. It is important to note that this result has been obtained without any adjustment to the values of the parameters describing the efficiency of surface magnetic braking or internal AM transport. This implies that the observed rotation rate of Kepler-444-A is perfectly in line with the rotational properties expected for such an old K0-type star. We thus find that Kepler-444-A does not exhibit a peculiarly fast rotation compared to similar old stars, which suggests that there is no reason for this star to be especially active. This is an important result regarding the physical nature of the Ly-$\rm \alpha$ variations observed for this star.

The degeneracy in the initial rotation of the star due to the braking of the surface by magnetised winds makes tracing back the rotational history of Kepler-444-A difficult which, therefore, can only be better constrained by looking at the indirect effects of rotation on the planetary system evolution. For example, the impact of dynamical and/or equilibrium tides on the architecture of the system is strongly related to the rotational properties of the host star, as well as the emission of high-energy luminosity and its effect on atmospheric escape. In the bottom panel of Fig. \ref{rotational_history}, the evolution of the EUV (solid line) and X-ray luminosity (dotted line) is shown, relative to the three initial surface rotations considered. These X-ray and EUV luminosities were computed according to Eqs.~\ref{eq:Tu2015} and \ref{eq:SF} (unless explicitly mentioned, the X-ray and EUV luminosities considered in the present study correspond to the ones predicted by these equations). During the early stages of the evolution, all rotators are in the saturated regime with emitted luminosities following the same trend. The higher the $\rm \Omega_{ini}$ value is, the longer the duration of the saturated regime. Therefore, the slow rotating case is the first one to switch to unsaturated conditions at $\rm \sim 125 ~ Myr$, followed by the medium and the fast ones. In the case of a fast initial rotation, the planets of the system would be exposed to high-energy irradiation for a rather prolonged period due to the longer duration of the saturated regime.

\subsection{Orbital evolution: Kepler-444-d and Kepler-444-e}

\begin{figure}
\centering
\subfigure{\includegraphics[width=0.45\textwidth]{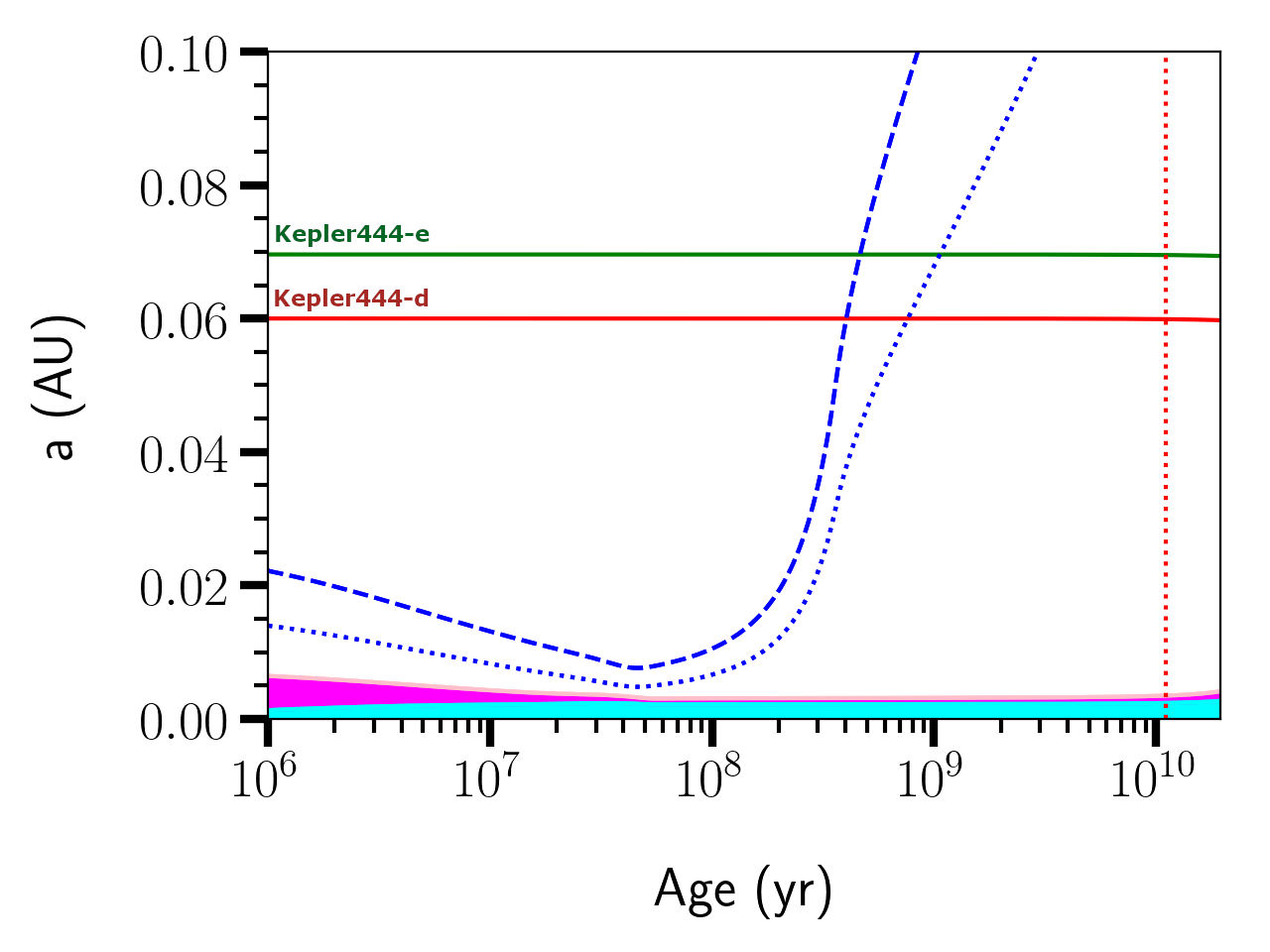}} 
\subfigure{\includegraphics[width=0.45\textwidth]{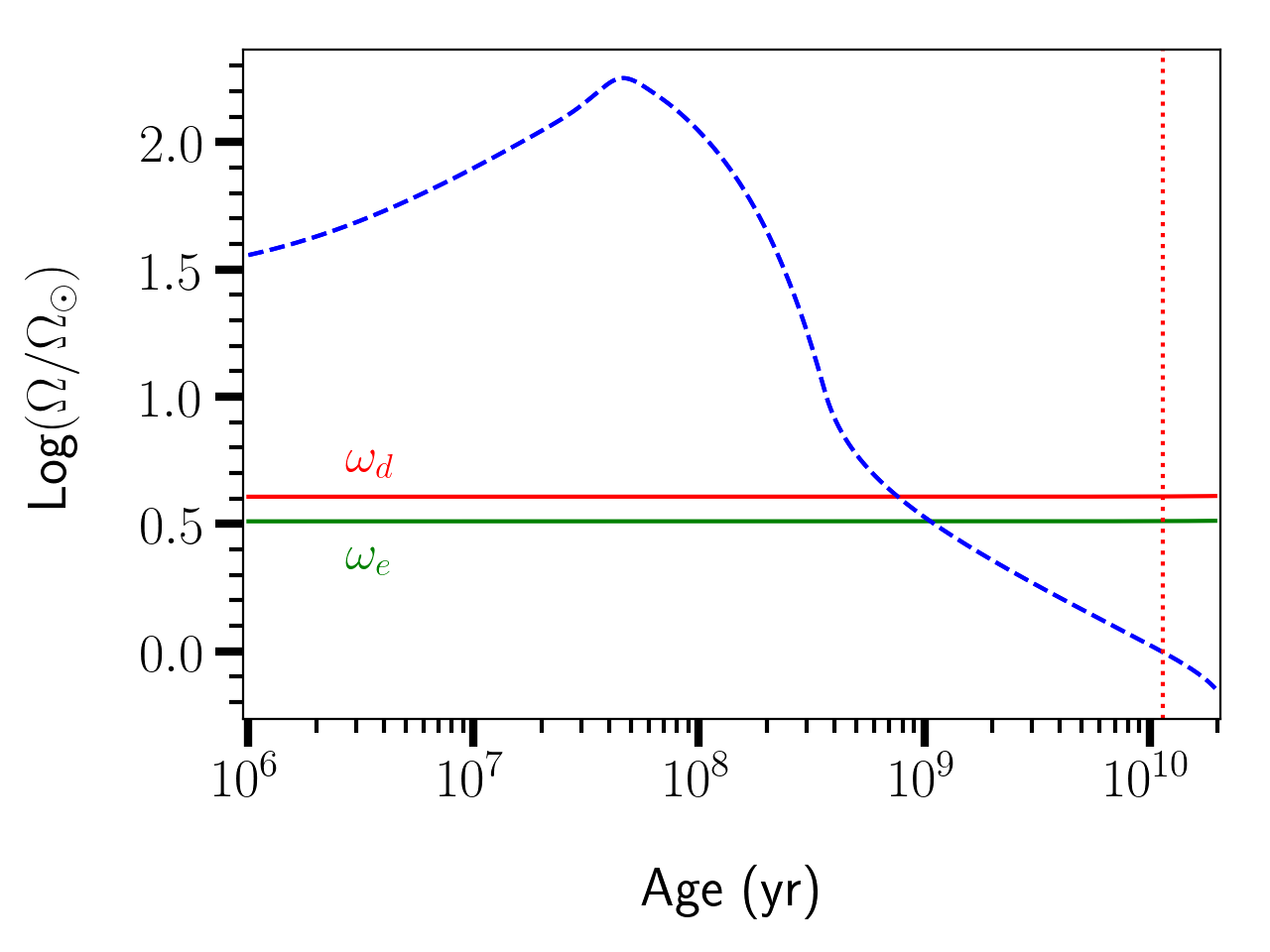}}
\caption{\footnotesize{\emph{Top panel:} Orbital evolution of Kepler-444-d and e (solid red and green line, respectively), starting with the current value of the orbital distance ($\rm a_{ini_{d}} = 0.06$, $\rm a_{ini_{e}} = 0.0696$), at a fixed mass ($\rm M_{d} = 0.1016~ M_{\oplus}$, $\rm M_{e} = 0.0921 ~M_{\oplus}$) and host star initial angular rotation $\rm \Omega_{in} = 18 \times \Omega_{\odot}$. The vertical red-dotted line corresponds to the age of the system ($\rm age = 11 ~ Gyr$). The blue-dashed lines indicate the evolution of the corotation radii, the blue-dotted ones indicate the evolution of the minimum initial orbital distances below which dynamical tides are no longer active. The magenta area represents the extension of the stellar convective envelope, while the cyan area represents the stellar radiative core. \emph{Bottom panel:} Orbital angular velocity of the planets and the double of the host star angular velocity (blue-dashed line) as a function of time. Dynamical tides are active as long as the orbital frequency of the planets is lower than double the host star surface angular rotation.}}
\label{Kep_DE}
\end{figure}

We studied the impact of dynamical tides on the total change of the orbital distance for the two planets for which an estimate of the mass is available, namely Kepler-444-d and Kepler-444-e. We first chose the current measured values of the orbital distances as initial input for our computations (see Table \ref{Planets_param}) and considered the masses to remain constant along the entire evolution. For the planetary masses, we took the value at the errorbar upper limit ($\rm M_{d} = 0.1016$ and $\rm M_{e} = 0.0921$ $\rm M_{\oplus}$), while for the host star we considered the case of a fast rotator ($\rm \Omega_{ini} = 18 \times \Omega_{\odot}$) in order to maximise the impact of dynamical tides. As a result of the simulation, we did not find any significant change in the orbital evolution of the system related to dynamical tides (see Fig. \ref{Kep_DE}). This result is due to the low mass ratio between planets and the host star ($\rm q = m_{pl}/M_{\odot}$), which substantially weakens the impact of dynamical tides. Moreover, the initial orbital distances of such low mass planets are too large for dynamical tides to have an appreciable impact on the evolution of the architecture of the system.

\begin{center}
\begin{table}
\begin{adjustbox}{max width=0.5\textwidth}
\begin{tabular}{cccc}
\hline \hline
Planet & $\rm a (AU)$ & $\rm R_{p}/R_{\oplus}$ & $\rm M_{p}/M_{\oplus}$ \\
\hline
\\
$\rm Kepler-444-b$ &   $\rm 0.04178^{+0.00079}_{-0.00079}$    &$\rm 0.403^{+0.016}_{-0.014}$ & - \\
\\
$\rm Kepler-444-d$ &  $\rm 0.0600^{+0.0011}_{-0.0011}$   &$\rm 0.543^{+ 0.018}_{-0.018}$   & $\rm 0.0364^{+0.0652}_{-0.0203}$  \\
\\
$\rm Kepler-444-e$ &   $\rm 0.0696^{+0.0013}_{-0.0013}$  &$\rm 0.559^{+ 0.017}_{-0.017}$   & $\rm 0.0336^{+0.0585}_{-0.0186}$\\
\\
\hline
\end{tabular}  
\end{adjustbox}
\captionof{table}{\footnotesize{Orbital and physical parameters for Kepler-444-b, Kepler-444-d, and Kepler-444-e. The orbital distances and the radius of Kepler-444-b are taken from \cite{Campante2015}, while the radii and masses of Kepler-444-d and e are taken from Paper I.}}
\label{Planets_param} 
\end{table}
\end{center}

\begin{figure}
\includegraphics[width=\linewidth]{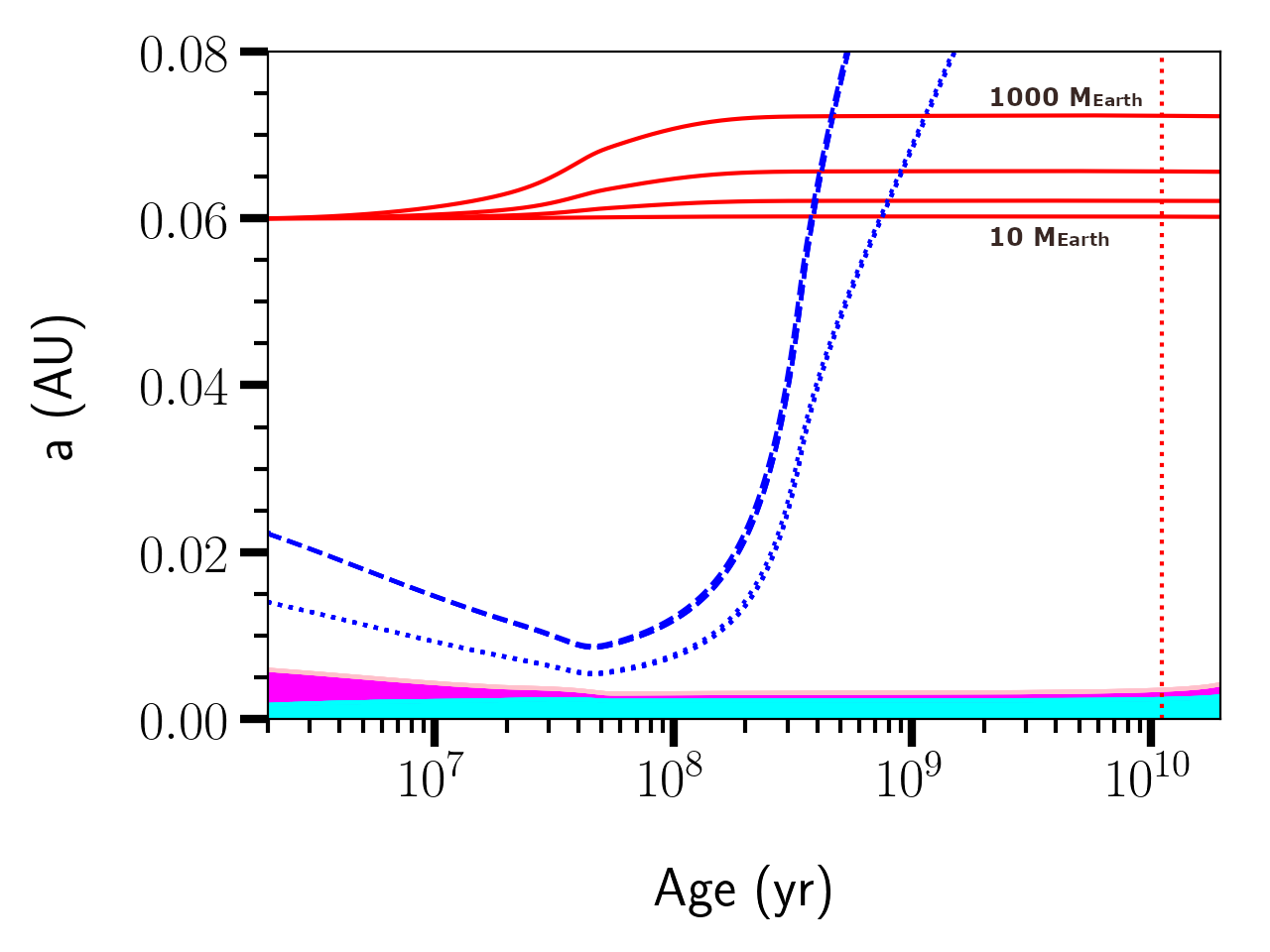}
\caption{\small{Impact of a change in the planetary mass on the orbital evolution for a planet located at the same distance as Kepler-444-d (initial value fixed to $\rm a_{ini_{d}} = 0.06$ AU). The four red lines span a range of masses from 10 to 1000 M$_{\oplus}$, with larger masses corresponding to more significant increases in the distances. The dashed lines, dotted lines, and shaded regions are the same as in Fig.~\ref{Kep_DE} (\emph{Top panel}).}}
\label{orb_mass}
\end{figure}

To investigate these two points in more detail, additional computations were first performed for the evolution of a planet with the same initial distance as Kepler-444-d (i.e. 0.06 AU), but for the following different values of the planetary mass: 10, 100, 317, and 1000 M$_{\oplus}$. The results of these computations are shown in Fig.~\ref{orb_mass}. Even in the case of a planet of 10 M$_{\oplus}$, no change in the orbital distance is seen. When the mass is increased to 100 M$_{\oplus}$, a slight increase in the orbital distance is obtained as a result of the more efficient dynamical tides. A significant impact of dynamical tides on the evolution of the semi-major axis is only seen for more massive planets and, in particular, for the 1000 M$_{\oplus}$ planet. This explains the evolution at a constant orbital distance obtained for the light planets of the Kepler-444 system seen in Fig.~\ref{Kep_DE}.

\begin{figure}
\includegraphics[width=\linewidth]{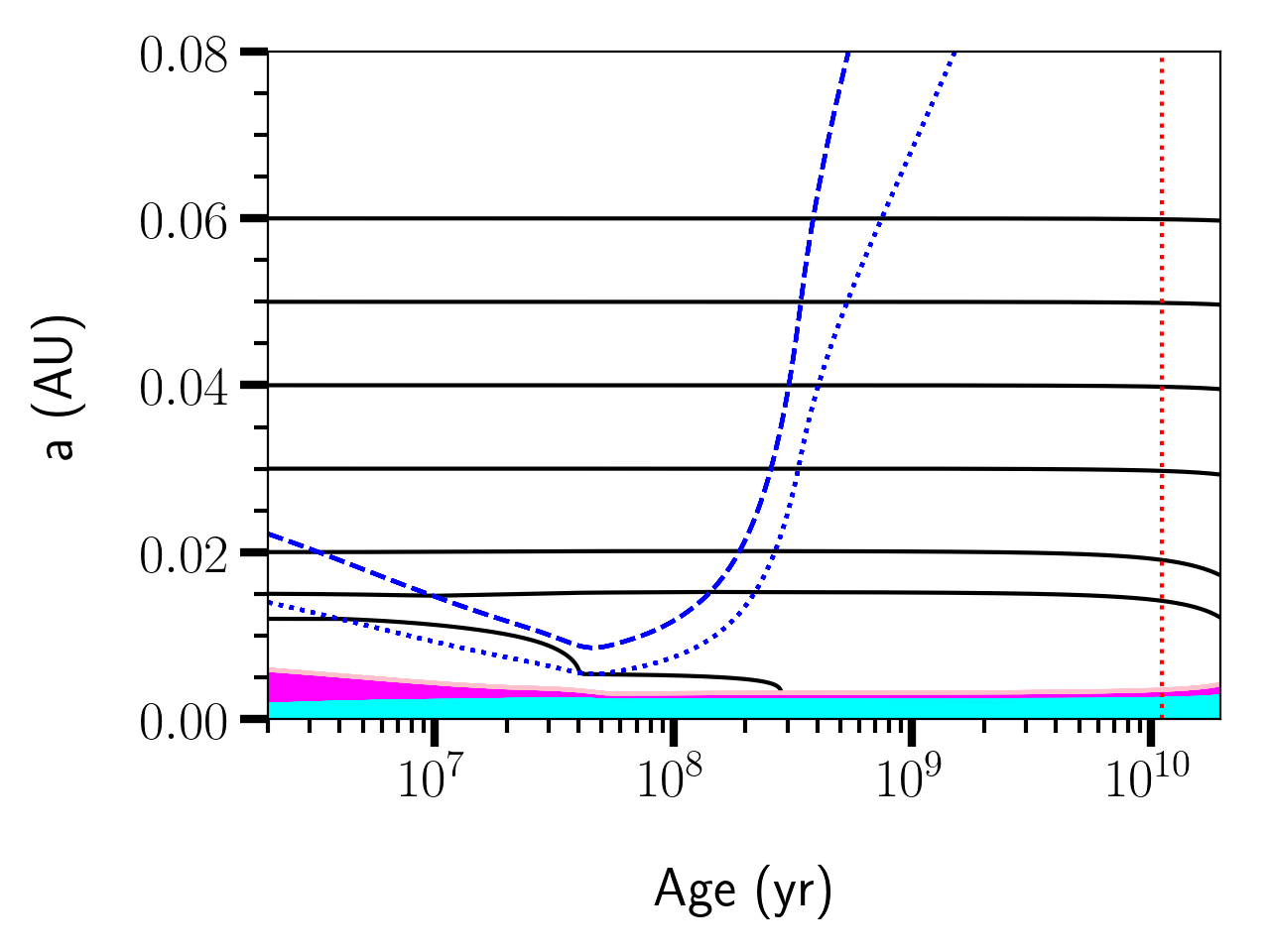}
\caption{\small{Impact of a change in the initial orbital distance for a planet with the maximal value of the mass allowed for Kepler-444-d (mass of the planet fixed to 0.1016 M$_{\oplus}$). The dashed lines, dotted lines, and shaded regions are the same as in Fig.~\ref{Kep_DE} (\emph{Top panel}).}}
\label{orb_distance}
\end{figure}

In the same way, the impact of the initial orbital distance on the efficiency of tides is studied by computing models with a mass fixed to the maximal value allowed for Kepler-444-d (0.1016 M$_{\oplus}$). Figure~\ref{orb_distance} shows the orbital evolution corresponding to different values of the initial semi-major axis. We find that only for values of the initial orbital distance lower than about 0.012 AU, a significant change is observed for the orbital evolution. This is again  a much lower value than the current orbital distance of the planets in the Kepler-444 system. This illustrates the fact that dynamical tides can only have an impact on very close-in and massive planets, which is not the case for the Kepler-444 system.

\subsection{Orbital evolution: Kepler-444-b}

We finally investigated the specific case of the innermost planet of the system (Kepler-444-b). As discussed above, the impact of dynamical tides can be more significant due to the closer orbit of Kepler-444-d to the host star. Although there is no mass measurement for Kepler-444-b, it is possible to have an estimate of this quantity using the radius-mass scaling law obtained by \citet{Valencia2006} for super-Mercury planets ($\rm R_{pl} \propto M_{pl}^{0.3}$). The mass of Kepler-444-b computed by means of this relation is about $\rm m_{b} = 0.048 \, M_{\oplus}$. As shown in Fig.~\ref{orb_distance}, no change in the orbit is expected for such a light planet at the initial orbital distance of Kepler-444-b (about 0.04 AU). This is confirmed by the specific computation of the orbital evolution of this planet using the method described in the previous section (as in the case of Kepler-444-d and Kepler-444-e), according to which the impact of dynamical tides on the planet orbit is also negligible in the case of Kepler-444-b, even considering an initial fast rotation for the host star.

\section{Impact of evaporation}

In the previous sections, we studied the orbital evolution of Kepler-444-d, Kepler-444-e, and Kepler-444-b using the current measurements for the orbital distances and planetary masses as initial inputs and testing the impact of different initial rotations of the host star. We did not find any appreciable changes in the planetary orbits. This result is mainly due to the low values of the planetary masses, or more precisely to the low values of the mass ratio $\rm q = M_{pl}/M_{\star}$ and the relatively large values for the initial orbital distances. In the preceding studies, we however made the simplifying assumption of a constant planetary mass along the whole evolution.

If we want to have a more comprehensive study of the system, we should include the impact of atmospheric evaporation. This process may be extremely relevant in changing the fate of the system, especially in the present case of low-mass close-in planets ($\rm a \leq 0.1 AU$) that are strongly exposed to high-energy stellar radiation, eventually undergoing atmospheric hydrodynamical escape \citep[e.g.][]{Ehrenreich2015, Jin2018}. The atmospheric escape may notably influence the evolution of the planetary radius and mass, with distinct efficiencies depending on multiple parameters related to both the planet properties and the rotational evolution of the host star. 
According to \cite{Jin2018}, low-mass low-density planets are most likely to entirely lose their envelopes on relatively short timescales ($\rm 2 ~ Myr$) due to their low gravitational binding energy. 
In case the planets formed outside of the ice line, they could have retained a significant water ice content while migrating towards the host star. This ice, which is heated up by the stellar radiation, can replenish the thin planetary atmosphere with H during the long evolution of the system. This would allow one to explain why, despite its old age, there would still be a phenomenon of hydrogen escape from these low-mass planets.

Using the Space Telescope Imaging Spectrograph of the Hubble Space Telescope, \citet{Bourrier2017} observed variations in the HI Lyman-$\rm \alpha$ line ($\rm 1215.6701 ~ \AA$) of Kepler-444-A. They detected significant variations at different observational epochs. These variations are up to the $\rm \sim 20\%$ during the transit of Kepler-444-e and Kepler-444-f. When no transit occurs, the variation is even stronger, up to the $\rm \sim40\%$ \citep{Bourrier2017}. The nature of these variations, which were observed during in and out of transit epochs of the two planets, could be attributed to stellar variability. However, their amplitude over short timescales ($\rm \sim 2-3$ hours) seems to be too intense for such an old main sequence K0-type star \citep{Bourrier2017}. Another interesting mechanism eligible to explain these variations is the presence of neutral H-rich exospheres trailing the Kepler-444 planets. According to this hypothesis, the authors attribute the $\rm 20\%$ variation to the transit of Kepler-444-e and Kepler-444-f, while the $\rm 40\%$ variation is attributed to the partial transit of an undetected outermost planet as of yet, Kepler-444-g. They performed simulations with the EVaporating Exoplanets (EVE) code to interpret the Ly-$\rm \alpha$ variations in terms of exospheric absorption, finding that these observations could be explained by neutral hydrogen mass loss rates of the order  of $\rm 10^{7} - 10^{8} ~g~s^{-1}$ for Kepler-444-d, e, and f. More observations of the systems would be necessary to eventually corroborate or definitely discard the hypothesis of the presence of H-rich exospheres and of a sixth planet Kepler-444-g. Eventual detections of hydrogen exospheres would give some hints as to the possibility of these planets being formed with a Ganymede-like composition and evolving while retaining a substantial amount of water-ice.

As discussed in Sect.~\ref{rothist}, our detailed modelling of the rotational evolution of Kepler-444-A does provide support to the hypothesis for which the HI-Ly$\rm \alpha$ variations are not related to stellar activity, showing that this star is not rotating particularly fast and that its surface angular velocity is perfectly in line with what is expected for the rotation rate of similar old stars. In the following, we aim to investigate whether the observed signal can be interpreted as due to the absorption by neutral H-rich exospheres arising from the dissociation of water molecules from oceans at the surface of the planet, under the assumption of being formed with a Ganymede-like composition \citep{Bourrier2017}.

In the next section, we thus study the impact of atmospheric evaporation on the evolution of the Kepler-444-e planet and compare our estimate for the mass loss rate and total mass loss to the ones reported in \citet{Bourrier2017}. We focus our study specifically on this planet because we can benefit from the availability of the planetary mass measurement and mass loss rate estimations as constraints to our computations.

\subsection{Kepler-444-e evaporation}

We computed a model for the orbital evolution of Kepler-444-e, this time accounting for atmospheric evaporation. We used the prescription for planetary atmospheres mass loss in the energy-limited regime configuration \citep{Lecavelier2007, Erkaev2007} as in the following:

\begin{equation}
\rm \dot{M}_{tot} = \eta \left( \dfrac{R_{XUV}}{R_{p}}\right)^{2} \frac{3}{4} \dfrac{F_{XUV} (sma)}{G \varrho_{pl} K_{tide}},
\end{equation}
\label{evap_eq}

where the factor $\rm R_{XUV}/R_{p}$ accounts for the planet cross-section increase due to EUV radiation, $\rm K_{tide}$ accounts for the enhancement in the mass loss rate of close-in planets, due to the proximity of the Roche-lobe surface to the planetary surface \citep{Erkaev2007}, $\rm F_{XUV}$ is the high-energy flux emitted by the host star, $\rm \varrho_{pl}$ is the average density of the planet, and $\rm G$ is the gravitational constant. The factor $\rm \eta$ represents the heating efficiency, namely the percentage of stellar energy used to heat up the planetary atmospheric layers. In \citet{Bourrier2017}, the estimate for the current atmospheric mass loss is $\rm \dot{M}^{tot}_{e} = \eta ~ 1.7 \times 10^9$ $\rm g~ s^{-1}$, which is calculated by taking into account their derived values for the current X-ray and EUV stellar fluxes ($\rm F_{X} (5-100 ~ \AA) \sim 0.1 $, $\rm F_{EUV} (100-912 ~ \AA) = 1.3^{+2.6}_{-1.1}$, in units of ($\rm erg~cm^{-2}~s^{-1}$), given at $\rm 1$ AU) and setting the factors $\rm R_{XUV}/R_{p}$  and $\rm K_{tide}$ to unity. They assumed that these fluxes remain constant along the whole evolution. They made the hypothesis that the planet formed beyond the ice line, with a Ganymede-like composition, for which the mean density is $\rm 1.94 ~ g ~ cm^{-3}$ and half of the planetary mass is made of water ice. Using a conservative heating efficiency ($\rm \eta = 0.1$), they estimated Kepler-444-e to have lost approximately $\rm 27\%$ of its initial water content after $\rm \sim 11$ Gyr. 

In the following, we present our results concerning the estimation of the Kepler-444-e atmospheric mass loss.\ This time, they were computed  by considering the evolution of the stellar rotation and the related emitted XUV flux.

\subsection{Impact of the rotational history}
\label{Rot_XUV}

The hypothesis of a constant XUV flux along the evolution of the system, as described in the previous section, does not provide a realistic scenario characterising the high-energy emission of the host star since the XUV luminosity may vary over a range of several orders of magnitude along its evolution, as shown in Fig. \ref{rotational_history}. In this context, it appears fundamental to consider the rotational history of the star and the related evolution of XUV luminosity in order to have a more realistic estimation of the planetary mass loss \citep[see e.g.][]{Johnstone2015}.

As a starting point, we used the estimation of the initial planetary mass given in \citet{Bourrier2017},  namely $\rm M_{pl_{in}} = 0.04344 ~ M_{\oplus}$\footnote{We recovered this value by integrating backwards in time the estimation of the atmospheric mass loss rate of Kepler-444-e computed in \citet{Bourrier2017} by using the energy-limited regime prescription, under the assumption of $\rm K_{tide} = 1$ and $\rm R_{XUV}/R_{p} = 1$, and for a heating efficiency of $\rm \eta = 0.1$.}, computed in the hypothesis of a constant XUV flux. Considering this value as the initial mass for Kepler-444-e, we tested whether by including an evolving XUV flux (linked to the rotational history of the host star) the planet would have survived to complete evaporation\footnote{The complete evaporation considered here corresponds to the ideal case in which the mass of the planet is completely eroded by the impact of XUV luminosity. In a more realistic scenario, it would be required to consider the eventual presence of a rocky core, which however is not completely certain in the case of Kepler-444-e.} and if not, what would have been the timescale needed to remove the totality of its mass. We thus took into account the evolution of the stellar XUV flux as predicted by our modelling of the evolution of the system using Eqs.~\ref{eq:Tu2015} and \ref{eq:SF}. In all these computations, the heating efficiency ($\rm \eta$) was fixed to 0.1. Since the rotational history of the host star is unknown, we decided to first consider the case of a slow rotator ($\rm \Omega_{in} = 3.2~\Omega_{\odot}$) in order to test the impact of the related high-energy flux emission. As a result, we determined that in this case the planet would be completely evaporated after $\rm \sim 50$ Myr. This result shows how the inclusion of an evolving XUV leads to much stronger lifetime mass losses, even in the case of a slow rotator\footnote{We can expect this evaporation timescale to become shorter for higher initial surface rotation rates, given that a star remains in the saturation regime longer for higher values of $\rm \Omega_{in}$.}. According to such a short timescale, it looks unlikely that the planet could have retained any atmosphere after $\rm \sim 11$ Gyr of evolution for an initial mass of $\rm 0.04344~M_{\oplus}$. \\

We may suppose that Kepler-444-e formed with an initial mass larger than $\rm 0.04344~M_{\oplus}$. This mass indeed was estimated integrating backwards in time the mass loss rates derived in the assumption of a constant XUV flux. Such an assumption entails an underestimation of the effective XUV flux and the related mass loss rates, especially at the early stages of the evolution.

\begin{center}
\begin{figure}
\includegraphics[width=\linewidth]{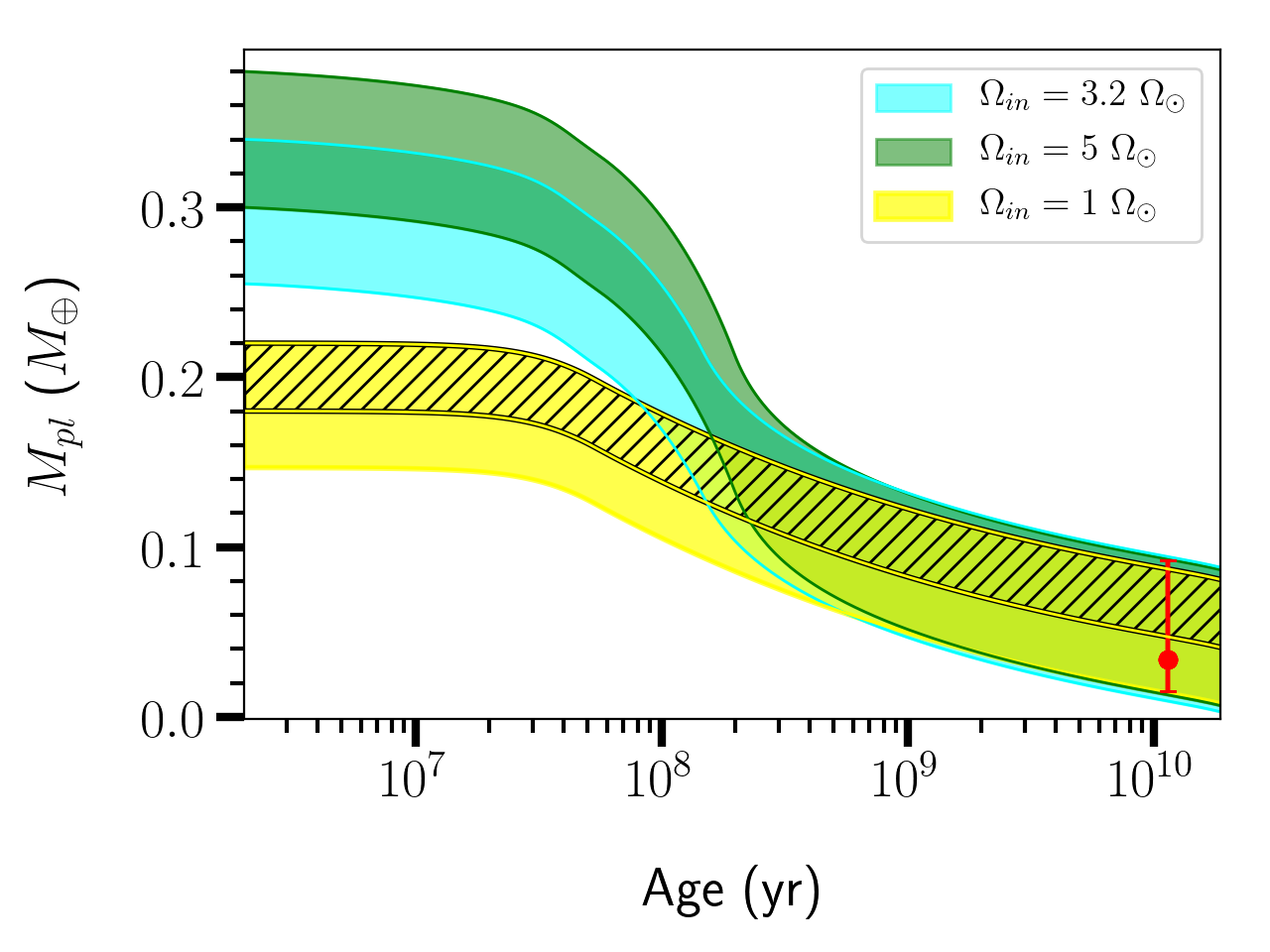}
\caption{\small{Evolution of the planetary mass estimated by considering different rotational histories for $\rm \eta = 0.1$ and $\rm R_{XUV}/R_{pl} = 1$. The yellow shaded area shows the evolution for $\rm \Omega_{in} = 1 ~\Omega_{\odot}$, the cyan shaded area for $\rm \Omega_{in} = 3.2 ~ \Omega_{\odot}$, and the green shaded area for $\rm \Omega_{in} = 5 ~ \Omega_{\odot}$. The black hatched area shows the evolution for $\rm \Omega_{in} = \Omega_{\odot}$, which is the range of initial masses compatible with an initial Ganymede-like composition.}}
\label{Mevap_omegas}
\end{figure}
\end{center}

This time, we integrated backwards in time the mass loss rates obtained in the case of a slow rotator and looked for the range of initial masses that reproduce the mass of the planet at the age of the system ($\rm \sim 11$ Gyr). We found that this range is $\rm \left(0.255 \sim 0.34\right)~M_{\oplus}$ (see Fig.~\ref{Mevap_omegas}). At this point, we tested whether this range of initial masses was compatible with the assumption that Kepler-444-e formed with an initial Ganymede-like composition, namely as a differentiated planet with a $\rm 50\%$ water-ice content. In a recent study, \citet{Michel2020} found that the water mass fraction of solid planets that formed beyond the ice line, orbiting a thick-disc population's host star, represents $\rm \sim 70\%$ of the total planetary bulk composition. We may consider this percentage as representative of the maximum fraction of the total mass loss allowed for Kepler-444-e, with respect to the initial mass.\\
We thus compared the range of initial masses found in the case of a slow rotator with the present mass of Kepler-444-e at the extremes of the errorbar (namely at $\rm 0.015$ and $\rm 0.0921~M_{\oplus}$), and we found that it would correspond to mass losses of $\rm 94\%$ and $\rm 73\%$ over the lifetime of the planet, with respect to the initial mass. The value that reproduces a final mass of $\rm 0.0336~M_{\oplus}$ within this range of initial masses is $\rm M_{pl_{in}} = 0.28~M_{\oplus}$, which leads to a percentage of total mass loss of $\rm 88\%$.\\
This result shows that in the case of a slow rotator, we predict initial masses that are not compatible with the assumption of an initial Ganymede-like composition. Indeed, we found total mass losses that are globally above the maximum percentage ($\rm > 70\%~M_{in_{pl}}$) allowed for such a planet. We would expect to obtain even stronger mass losses and consequently to predict larger initial masses in the case of moderate ($\rm \Omega_{in} = 5~\Omega_{\odot}$) and fast ($\rm \Omega_{in} = 18~\Omega_{\odot}$) rotators, given that a star remains in the saturation regime of emission longer for higher values of $\rm \Omega_{in}$. According to our results, we disfavour the hypothesis according to which Kepler-444-e has retained a fraction of its initial water-ice content after $\rm \sim 11$ Gyr, due to the strong mass losses experienced by the planet, even in the case of a slow-rotating host star.

\begin{figure}[t]
\centering
\includegraphics[width=0.4\textwidth]{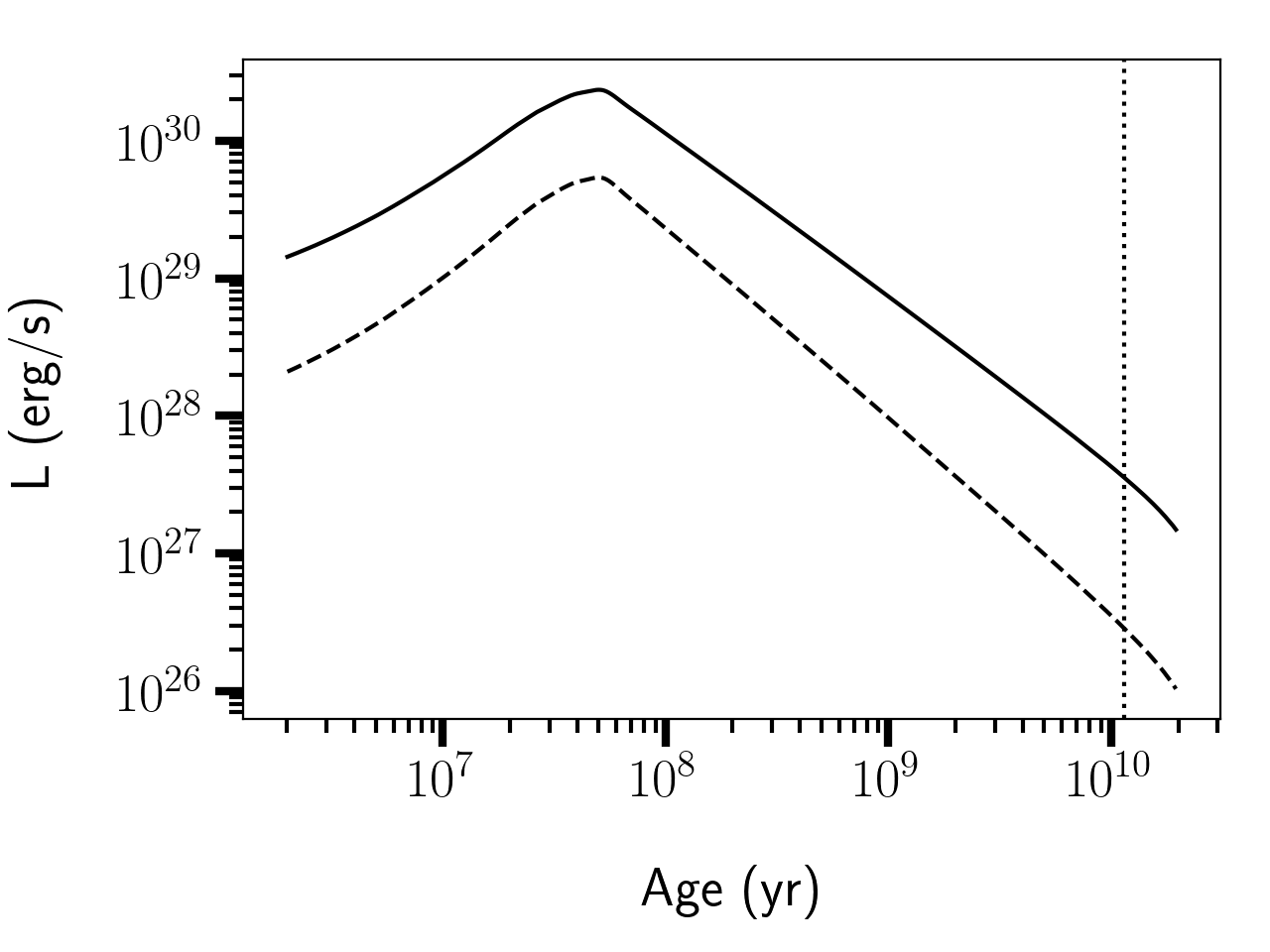}
\caption{\small{Evolution of the EUV and X-ray luminosity (solid and dashed line, respectively) computed using Eqs.~\ref{eq:Tu2015} and \ref{eq:SF}} for the super-slow rotating case ($\rm \Omega_{in} = \Omega_{\odot}$). The vertical line indicates the age of the system.}
\label{Kep_rotation_omega_sun}
\end{figure}

In this context, we are interested in exploring how much the lifetime mass loss of Kepler-444-e may change when considering the host star has an initial rotation rate smaller than $\rm 3.2~\Omega_{\odot}$. We may assume that the host star had an initial surface rotation rate equal to the one of the present Sun ($\rm \Omega_{in} = \Omega_{\odot}$). Such a star is a super-slow rotator. Statistically, super-slow rotators appear to be much rarer than slow, medium, and fast rotators looking at the observed distribution of surface rotations for stars in star-forming regions and young open clusters. Based on the study by \citet{Gallet2015}, we indeed note that no young stars (i.e. with an age less than 4 Myr) in the mass range of $\rm 0.7 - 0.9~M_{\odot}$ are observed with a surface angular velocity lower than the initial value of $\rm 1~\Omega_{\odot}$ adopted for the super-slow rotating case, and that the mean value for the 25th rotational percentile is found to be about $\rm 3~\Omega_{\odot}$ for these stars. Nevertheless, it might still be interesting to estimate the mass loss rates of Kepler-444-e, assuming this peculiar stellar rotational history. In Fig.~\ref{Kep_rotation_omega_sun}, we show the evolution of the X-ray and EUV luminosity for a super-slow rotator. We notice that for such a low value for the initial surface rotation rate, the emission of high energy occurs in non-saturated conditions for the whole evolution of the system. In this case, we predict a range of initial masses of $\rm (0.147 \sim 0.22)~M_{\oplus}$ to which total mass losses of $\rm 89\%$ and $\rm 58\%$ of $\rm M_{pl_{in}}$ correspond when compared with the final masses of $\rm 0.015$ and $\rm 0.0921~M_{\oplus}$, respectively. We see that in this case, we predict an initial mass of the planet ($\rm 0.22~M_{\oplus}$) able to reproduce the present mass of Kepler-444-e at the upper limit of the errorbar ($\rm 0.0921~M_{\oplus}$) that is compatible with an initial Ganymede-like composition, namely we predict a percentage of the total mass loss below $\rm 70\%$. There is actually a restricted range of initial masses that provides the same kind of result for a corresponding range of final masses: For $\rm M_{in_{pl}} \in \left[0.18 , 0.22 \right] ~M_{\oplus}$, we reproduce $\rm M_{fin} \in \left [ 0.053, 0.0921 \right] ~M_{\oplus} $ and total mass losses below $\rm 70\%$.\\

In Fig.~\ref{Mevap_omegas}, we represent our predictions and evolution of the planetary mass for different rotational histories of the host star. In particular, we show the results obtained for the super-slow, slow, and medium rotators (yellow, cyan, and green shaded areas, respectively). For the super-slow rotator, we also indicate the range of initial masses that would be compatible with the initial composition assumed for Kepler-444-e (black hatched area). 


\subsection{Impact of XUV-luminosity prescriptions}
\label{rot_hist_XUV}

As mentioned in previous sections, the rotational history of a given star is unknown. Nevertheless, for stars hosting planetary systems, it may be possible to infer some indications about preferential rotational histories by studying their direct impact on the evolution of the planets \citep{Kubyshkina2019}. 
In the previous section, we found that in the case of a slow rotator ($\rm \Omega_{in} = 3.2~\Omega_{\odot}$), we predict initial masses for Kepler-444-e that are not compatible with an initial Ganymede-like composition. We thus considered the particular case of a super-slow rotator, finding a restricted range of initial masses ($\rm 0.18 \lesssim M_{in_{pl}}/M_{\oplus} \lesssim 0.22$) with which our results are compatible, with total mass losses below the $\rm 70\%$ threshold of the initial mass. According to this result, we might establish that the super-slow rotator case would constitute a preferential pathway for the rotational history of the host star.

However, there is another important physical ingredient intervening in the evaluation of the atmospheric mass loss that may change the estimation of the initial planetary mass significantly, namely the XUV-luminosity prescription. It is possible that the estimation of the planetary mass loss might vary considerably, depending on the XUV-luminosity prescription used, even if the same rotational history is considered for the host star. So far, we have used Eq.~\ref{eq:Tu2015} for the computation of the X-ray luminosity and Eq.~\ref{eq:SF} for the derivation of the corresponding EUV luminosity \citep{Wright2011,SanzForcada2011}. It is worth investigating how the results obtained for the evaporation of Kepler-444-e in the case of a super-slow rotator may change by using an alternative XUV-luminosity prescription.

\begin{figure}
\centering
\subfigure{\includegraphics[width=0.45\textwidth]{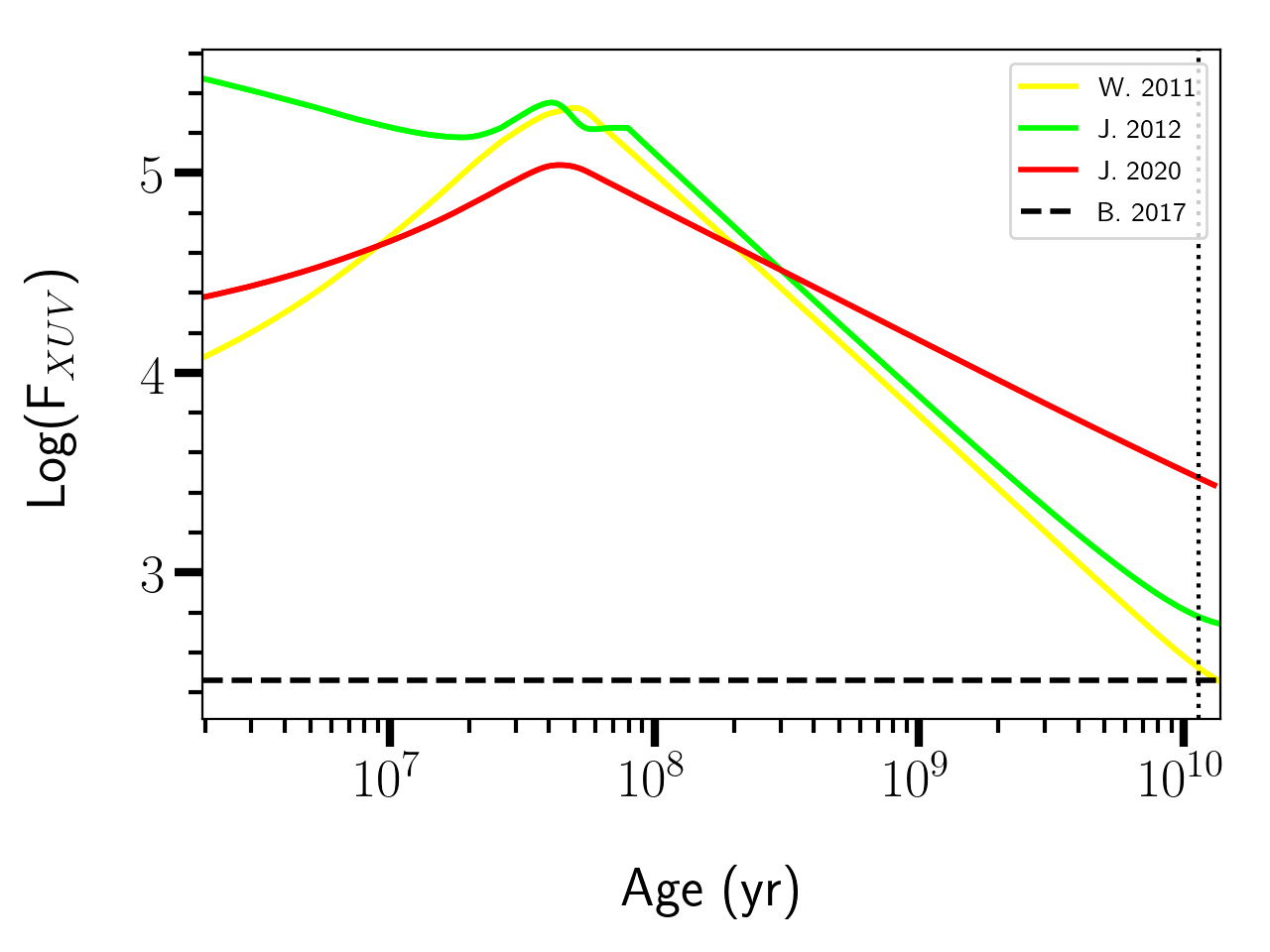}} 
\subfigure{\includegraphics[width=0.46\textwidth]{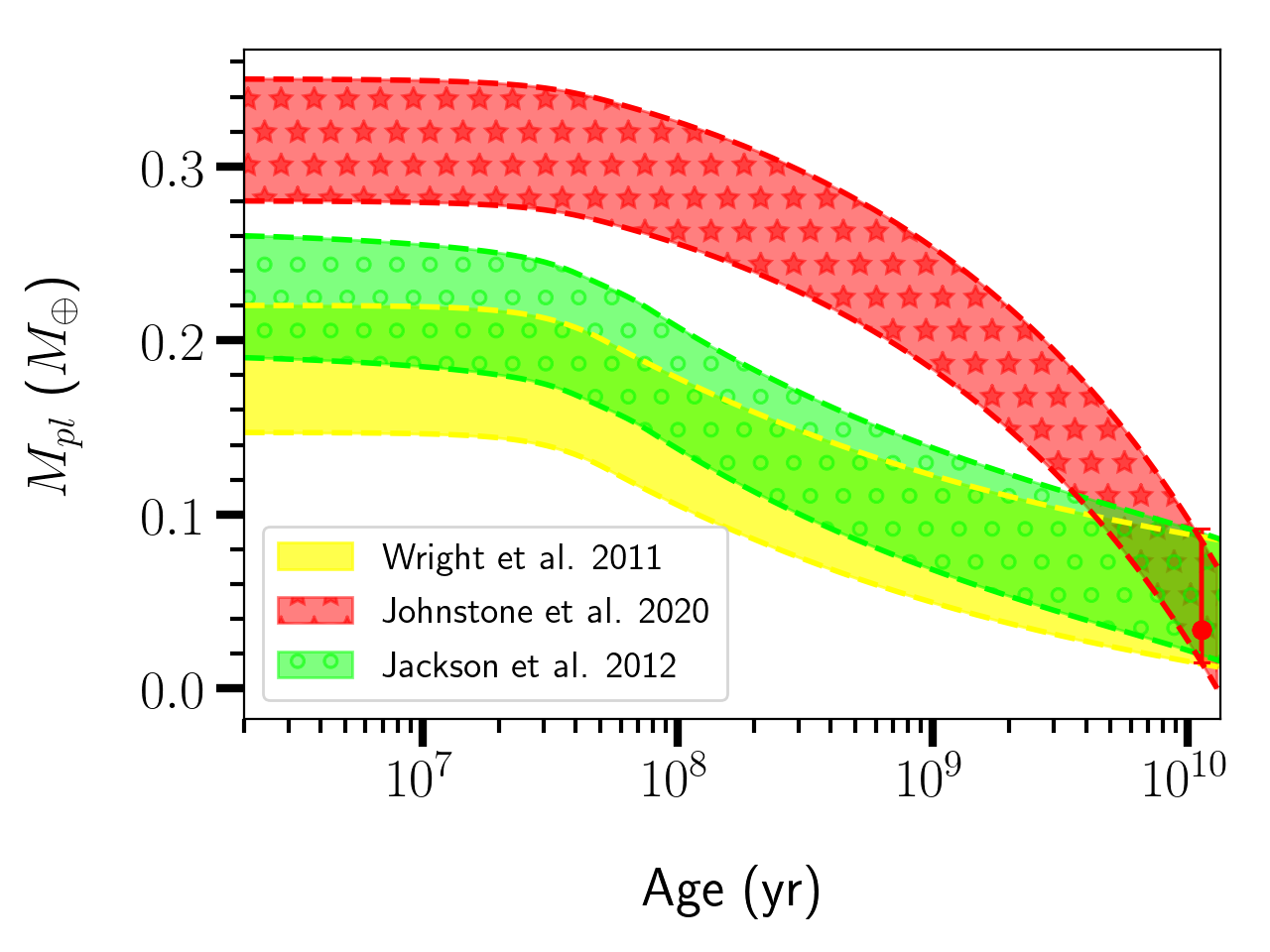}}
\caption{\small{\emph{Top panel:} Comparison among the evolutions of XUV fluxes computed by following the work of \citet{Wright2011} (solid yellow line), \citet{Jackson2012} (solid green line), and \citet{Johnstone2020} (solid red line). The horizontal black-dashed line represents the estimation of the XUV flux of \citet{Bourrier2017}.  \emph{Bottom panel:} Evolution of the planetary mass computed by considering different XUV-luminosity prescriptions. The colour code corresponds to the one described in the top panel.}}
\label{Fluxes_comp}
\end{figure}

For this purpose, following the work of \citet{Johnstone2020}, we adopted new prescriptions for the computation of the X-ray and EUV luminosities (see Eqs.~\ref{eq:J2020} and \ref{eq:J2020_  euv} in Sect.~\ref{sect:orbit}) and using the same physics set-up defined in Sec.~\ref{Rot_XUV}, we recomputed the evolution of the planetary mass. In the top panel of Fig.~\ref{Fluxes_comp}, a comparison between the evolution of the different XUV fluxes is presented, the solid red and yellow lines being the XUV flux computed by following \citet{Johnstone2020} and \citet{Wright2011,SanzForcada2011}, respectively, while the horizontal black-dashed line represents the estimation of the XUV flux at the age of the system as reported in \citet{Bourrier2017}. For the sake of completion, we also show the evolution of the XUV flux computed as in \citet{Jackson2012}, reminding us that in this case it does not directly depend on the rotational history of the host star. One of the main features emerging from this plot is that the slope of the red line is flatter compared to the one of the yellow line, especially in the unsaturated regime, leading to an estimation of the XUV flux at the age of the system that is almost one order of magnitude larger than the one indicated by the yellow line and by \citet{Bourrier2017}. As a consequence, we expect to get a much more efficient atmospheric mass loss when using the new XUV-luminosity prescription in the evaporation routine. Indeed, in this case, we predict that the initial mass of the planet should have ranged between roughly $\rm 0.28$ and $\rm 0.35~M_{\oplus}$ in order to be able to reproduce its mass at the age of the system. This range of initial masses implies total mass losses that are above the maximum percentage for being compatible with an initial Ganymede-like composition. 

In the light of this result, it seems that there is no chance for the planet to retain a fraction of its initial water-ice content with the new XUV-luminosity prescription (even in the case of a super-slow rotator) if we assume it formed with  a Ganymede-like composition, unless some physical mechanisms intervene reducing or inhibiting the mass loss.

\subsection{Impact of the effective XUV-absorption radius}
\label{sect:Rxuv}

It is worth noting that the estimations of the atmospheric mass loss presented so far rely on the assumption that the factors ($\rm \beta =  R_{XUV}/R_{pl}$) and ($\rm K_{tide}$) in Eq.~\ref{evap_eq} are set to unity, which is consistent with \citet{Bourrier2017}. However, accounting for their variations as a function of time might have a significant impact on the final results. In particular, accounting for the presence of an absorption radius located at larger distances with respect to the surface of the planet increases the effective area heated by the stellar high-energy radiation, consequently leading to a more efficient evaporation. 

Here, we consider the contribution of a factor of $\rm \beta > 1$ in the estimation of the planetary lifetime mass loss. For the computation of this factor, we refer readers to the work of \citet{Salz2016}.

In the light of the results obtained in the previous sections, we aim to study whether the particular set of conditions that we found to favour the hypothesis of an initial Ganymede-like composition for $\rm \beta =1$ are still valid for $\rm \beta > 1$. These conditions require that the host star evolved as a super-slow rotator, emitting XUV radiation as predicted by Eq.~\ref{eq:Tu2015} and \ref{eq:SF}, and that the planet had an initial mass in the range of $\rm (0.18 - 0.22)~M_{\oplus}$.

We thus recomputed the evaporation of the planetary mass, this time including the contribution of the effective XUV-absorption radius. The results related to these computations are shown in Fig.~\ref{Rxuv} (solid red lines). We notice that the impact of the $\rm R_{XUV}$ radius is significant, leading to much stronger evaporation rates that would remove the totality of the planetary mass within the first $\rm \sim 50$ Myr of evolution. Such strong mass losses are incompatible with the hypothesis that Kepler-444-e could have retained any atmosphere after 11 Gyr of evolution if it formed with a Ganymede-like composition. The simulations obtained for $\rm \beta > 1$ thus show that, even in the case of a super-slow rotating star, the evaporation rates are strong enough to efficiently remove a mass content as large as the mass of Kepler-444-e in relatively short timescales.

\subsection{Impact of the heating efficiency}
\label{sect:eta}

As a last point, we would like to consider the role of the heating efficiency. In our study, we treated the evaporation using a value for the heating efficiency of $\rm 10\%$ to compare our results with the ones of \citet{Bourrier2017}. \citet{Lopez2017} indicates that while this value is usually used for planets with solar composition atmospheres, in the case of higher metallicity atmospheres it should be significantly lower. This is due to the higher mean-molecular weight that would reduce the scaleheight in the evaporative wind and increase the number of scaleheights between the XUV photosphere and the sonic point; additionally, this is due to the fact that in this regime the metal atomic lines are dominant coolants \citep{Lopez2017}. Applying the scaling relation found by \citet{Ercolano2010} between heating efficiency and atmospheric metallicity ($\rm \eta \propto Z^{-0.77}$) to pure water atmospheres, \citet{Lopez2017} finds a value of $\rm \eta$ closer to $\rm 1\%$. The same scaling relation is also used in \citet{Mordasini2020}. In this work, the author points out that the results based on this scaling should be considered with some caution, recalling that the relation of \citet{Ercolano2010} is derived from studies of X-ray heating of protoplanetary disc atmospheres rather than planetary atmospheres. Moreover, he argues that while the evaporation rate seems to likely decrease with increasing metal content (given the dominant role of atomic lines as coolants), the exact dependencies have not been explored yet \citep{Mordasini2020}.

\begin{center}
\begin{figure}
\includegraphics[width=\linewidth]{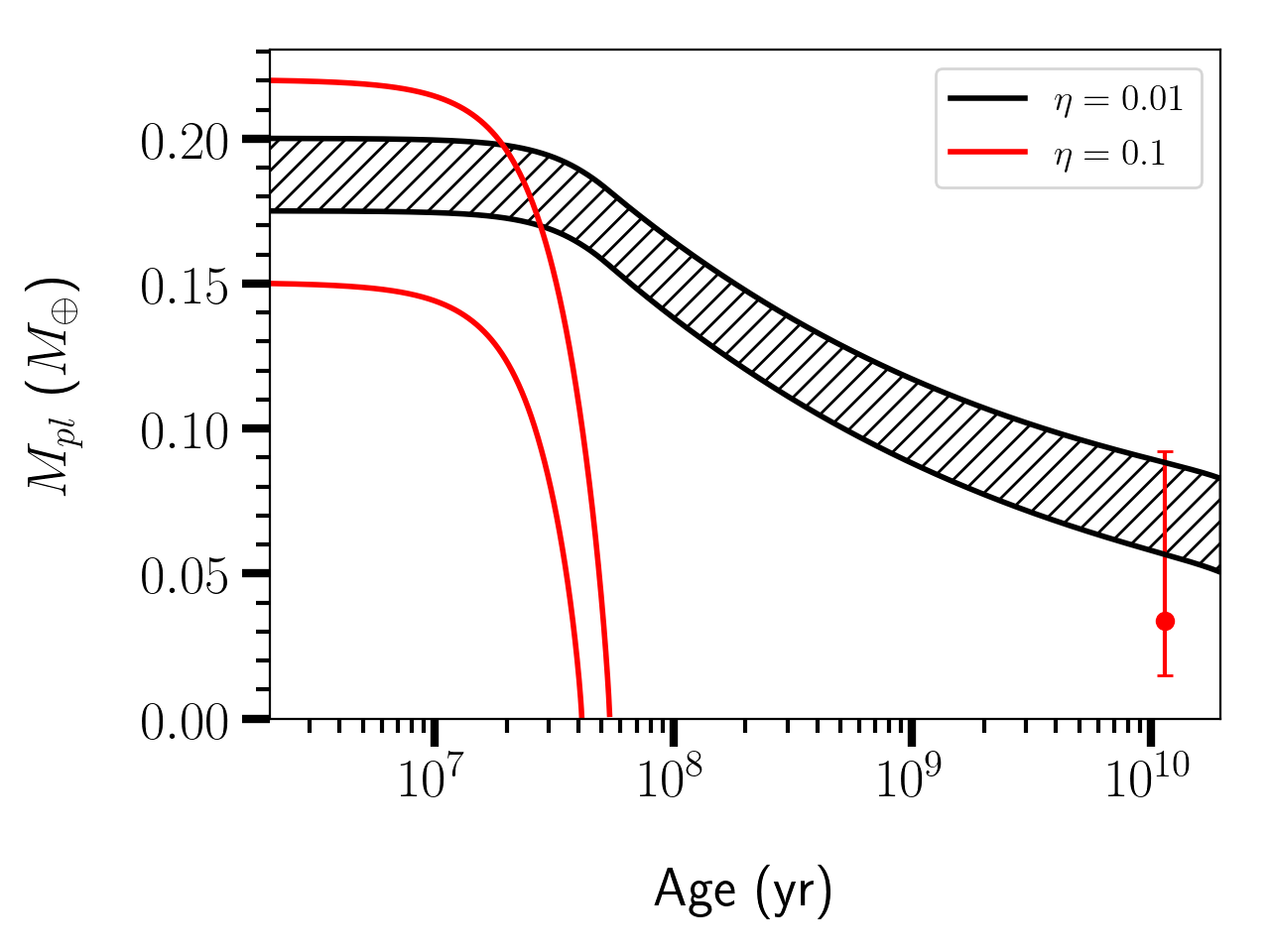}
\caption{\small{Evolution of the planetary mass computed with $\rm \beta > 1$, $\rm \Omega_{in} = \Omega_{\odot}$ and XUV formalism as in Eq.~\ref{eq:Tu2015} and \ref{eq:SF}. The solid red and black lines show the evolution for $\eta = 0.1$ and $\rm eta = 0.01$, respectively. The black shaded area indicates the range of initial masses compatible with a lifetime mass loss below $\rm 70\%~M_{in_{pl}}$.}}
\label{Rxuv}
\end{figure}
\end{center}

Given the uncertainties related to the value of the heating efficiency, we recomputed the evolution of the planetary mass for a slow and super-slow rotator with $\rm \beta > 1$, but this time for $\rm \eta = 0.01$. 
As a result, we found that in the case of the slow rotator, we predict initial masses that are not compatible with a Ganymede-like composition, even for such a low value of $\rm \eta$. Instead, for the super-slow rotator case, we found that the lower value of the heating efficiency rather compensates for the impact of $\rm R_{XUV}$, leading us to recover a range of initial masses ($\rm 0.175 \sim 0.20~M_{\oplus}$) that reproduce the mass of Kepler-444-e at the age of the system (see Fig.~\ref{Rxuv}, black hatched area) for a percentage of the total mass loss lower than $\rm 70\%$. \\

As stated before, some caution is required when using a heating efficiency as low as 0.01 in the case of metal rich atmospheres. Moreover, we recall that these simple evaporation computations tend to underestimate the mass loss rate, especially at the early stages of the evolution when the radius of the planet would be much more inflated \citep{Lopez2014,Chen2016,Zeng2016}, allowing for a larger effective absorption area of the stellar high-energy radiation together with a decrease in the gravitational well for escaping particles. More importantly, we note that if we compute the value of the restricted Jeans escape parameter\footnote{Defined as the Jeans escape parameter without accounting for Roche-lobe effects and hydrodynamic velocities, evaluated at $\rm R_{pl}$, considering the equilibrium temperature of the planet $\rm T_{eq}$ ($\rm \Lambda = G M_{pl} m_{H} / (k_{b} T_{eq} R_{pl})$)} of Kepler-444-e at the beginning of the evolution, assuming a maximal value for the initial mass $\rm M_{pl} = 0.22~M_{\oplus}$, a minimum value for the radius $\rm R_{pl} = 0.559~R_{\oplus}$, and a Bond albedo $\rm A = 0.5$, we obtain $\rm \Lambda \simeq 4$. For such a low value of the restricted Jeans escape parameter ($\rm \Lambda \lesssim 20$, \citealt{Fossati2017,Kubyshkina2018b}), the atmospheric escape occurs in the so called boil-off regime, in which the escape is driven by a combination of intrinsic thermal energy and low gravity. During this phase, the energy-limited formula has been shown to significantly underestimate the mass loss rates \citep{Lammer2016,OwenWu2016,Stokl2016,Fossati2017}. We thus see that, even in the most favourable case of a low heating efficiency of 0.01, the Ganymede hypothesis is disfavoured by evaporation computations for Kepler-444-e once the evolution of the stellar rotation and the related emitted XUV flux is taken into account.

\section{Conclusion}

In this paper we aimed to study the past history of the planets orbiting Kepler-444-A, using all the constraints available for this system.\\ We computed rotating stellar models of Kepler-444-A using the stellar parameters provided in Paper I and selecting three different values for the initial surface rotation rate, which are representative of initially slow, moderate, and fast rotators. We correctly reproduced the observed surface rotation rate of Kepler-444-A, finding that this star does not exhibit a particularly fast rotation compared to similar old stars. This result suggests that there is no reason for this star to be exceptionally active, favouring the hypothesis that the HI-Ly$\rm \alpha$ variations observed for this system are not related to stellar activity.\\

Coupling the rotating stellar models to our orbital evolution code, we studied the impact of dynamical and equilibrium tides on the evolution of Kepler-444-d and Kepler-444-e. Given the low mass ratio between planets and the host star and their relatively large orbital distances, we found that tides do not play a significant role in shaping the architecture of this system.\\

Afterwards, we studied the impact of atmospheric evaporation on Kepler-444-e. We investigated whether our estimated lifetime mass losses are compatible with the formation hypothesis of the planet with a Ganymede-like composition retaining a fraction of the initial water-ice content (representing up to the $\rm 70\%$ of the initial mass) at its current age, allowing the presence of a H-rich exosphere trailing the planet.

We computed XUV-flux evolutionary tracks for a slow rotator and estimated the mass loss rates using the energy-limited formalism. Starting with the initial planetary mass derived in \citet{Bourrier2017}, we found that our mass loss rates would remove the entire mass within the first $\rm 50$ Myr of evolution. Integrating the mass loss rates obtained for a slow rotator  backwards in time, we then predicted the range of initial masses that allowed us to reproduce the current mass of Kepler-444-e. However, we find total mass losses above the maximum percentage of the initial water-ice content allowed for a Ganymede-like planet. The same result applies to the case of medium and fast rotators to which larger total mass losses correspond.

We tested whether the same result holds for a star having a very low initial surface rotation rate (super-slow rotator). We found that for such a peculiar rotational history, it is possible to predict initial masses compatible with an initial Ganymede-like composition, but only for a particular set of conditions, requiring that the X-ray and EUV luminosities are computed following the works of \cite{Wright2011} and \cite{SanzForcada2011} and that the effective XUV-absorption radius is equal to the radius of the planet ($\rm \beta = 1$). This solution does not hold any more when using an alternative XUV-luminosity prescription, such as the one derived by following the work of \cite{Johnstone2020}; similarly, when including the impact of the effective XUV-absorption radius ($\rm \beta > 1$), we estimate total mass losses that are above the maximum percentage allowed for such a planet to be compatible with an initial Ganymede-like composition. Only when adopting a value of $\rm \eta$ as low as $\rm 1\%$, in the case of $\rm \beta >1$, we do retrieve a solution for the initial planetary mass compatible with our assumption on the initial composition, but only for a super-slow rotator and for XUV luminosities computed following \citet{Wright2011,SanzForcada2011}.

We note that the need to adopt an exceptionally super-slow rotating star and/or a very low value for the heating efficiency to be in agreement with the initial composition assumed for Kepler-444-e point towards disfavouring the hypothesis that the Ly-$\rm \alpha$ variations observed for this system are related with the presence of H-rich escaping atmospheres today, arising from the oceans at the planet surface. We recall that our rotating models also do not support a stellar origin for the observed HI-Ly$\rm \alpha$ variations, thus the origin of this detected variability remains an open question.

\section*{Acknowledgements}
We thank the referees for their useful comments that helped to improve the quality of the manuscript. This project has been supported by the Swiss National Science Foundation grant 200020-172505. G.B. acknowledges funding from the SNF AMBIZIONE grant No 185805 (Seismic inversions and modelling of transport processes in stars). P.E. and G.M. have received funding from the European Research Council (ERC) under the European Union's Horizon 2020 research and innovation program (grant agreement No 833925, project STAREX). V.B. acknowledges support by the Swiss National Science Foundation (SNSF) in the frame of the National Centre for Competence in Research “PlanetS”. This project has received funding from the European Research Council (ERC) under the European Union’s Horizon 2020 research and innovation programme (project Four Aces grant agreement No 724427; project Spice Dune grant agreement No 947634).

\bibliographystyle{aa}
\bibliography{biblioarticleKepler}

\end{document}